\documentclass[twocolumn]{aastex631}
\usepackage{amsmath}
\usepackage{multirow}

\begin{document}

\title{Cross-correlating dark sirens and galaxies: constraints on $H_0$ from GWTC-3 of LIGO-Virgo-KAGRA}
\author{Suvodip Mukherjee}\email{suvodip@tifr.res.in}
\affiliation{Department of Astronomy and Astrophysics, Tata Institute of Fundamental Research, Mumbai 400005, India}
\affiliation{Perimeter Institute for Theoretical Physics, 31 Caroline St. North, Waterloo, ON NL2 2Y5, Canada}
\author{Alex Krolewski}\email{akrolews@uwaterloo.ca}
\affiliation{AMTD Fellow, Waterloo Centre for Astrophysics, University of Waterloo, Waterloo ON N2L 3G1, Canada}
\affiliation{Perimeter Institute for Theoretical Physics, 31 Caroline St. North, Waterloo, ON NL2 2Y5, Canada}
\author{Benjamin D. Wandelt}\email{wandelt@iap.fr}
\affiliation{Institut d'Astrophysique de Paris, UMR 7095, CNRS, Sorbonne Universit\'e, 98bis Boulevard Arago, 75014 Paris, France}
\affiliation{Center for Computational Astrophysics, Flatiron Institute, 162 5th Avenue, 10010, New York, NY, USA}
\author{Joseph Silk}
\email{joseph.silk@physics.ox.ac.uk, silk@iap.fr}
\affiliation{Institut d'Astrophysique de Paris, UMR 7095, CNRS, Sorbonne Universit\'e, 98bis Boulevard Arago, 75014 Paris, France}
\affiliation{Beecroft Institute for Cosmology and Particle Astrophysics, University of Oxford, Keble Road, Oxford OX1 3RH, UK}
\affiliation{The Johns Hopkins University, Department of Physics \& Astronomy, Bloomberg Center for Physics and Astronomy, Room 366, 3400 N. Charles Street, Baltimore, MD 21218, USA}



\begin{abstract}
We apply the cross-correlation technique to infer  the Hubble constant ($H_0$) of the Universe using gravitational wave (GW) sources without electromagnetic counterparts (dark sirens) from the third GW
Transient Catalog (GWTC-3) and the photometric galaxy surveys 2MPZ and WISE-SuperCOSMOS, and combine these with the bright siren measurement of $H_0$ from GW170817. The posterior on $H_0$ with only dark sirens is uninformative due to the small number of well-localised GW sources. Using the eight well-localized dark sirens and the binary neutron star GW170817 with EM counterpart, we obtain a value of the Hubble constant $H_0= 75.4_{-6}^{+11}$ km/s/Mpc (median and 68.3$\%$ equal-tailed interval (ETI)) after marginalizing over the matter density and the GW bias parameters. This measurement is mostly driven by the bright siren measurement and any constraint from dark sirens is not statistically significant. In the future, with more well-localized GW events,  the constraints on expansion history will improve.
\end{abstract}



\section{Introduction} \label{sec:intro}

Discovery of gravitational waves (GW) \citep{PhysRevLett.116.061102} by the LIGO-Virgo-KAGRA (LVK) collaboration \citep{LIGOScientific:2014pky,Martynov:2016fzi,Acernese_2014,PhysRevLett.123.231108,Akutsu:2018axf,KAGRA:2020tym}  has opened a new observational window to study the cosmos using transient sources such as binary neutron stars (BNSs), binary black holes (BBHs) and neutron star black hole mergers (NSBHs).  GW sources are uniquely accurate tracers of the luminosity distance. They can therefore be used to measure the expansion history of the Universe (\citet{Schutz}). This fact has earned GW sources the name  \textit{standard sirens}. However, one of the key ingredients required to measure the expansion history using GW sources is an independent measurement (or inference) of the GW source redshifts.

In the absence of electromagnetic counterparts, a promising way to infer the GW source redshifts is through spatial cross-correlation of the GW sources with galaxies  {of known redshift. This is an application of the clustering redshift method} 
\citep{Newman:2008mb, Menard:2013aaa, Schmidt:2013sba} 
 {that has also been used for the calibration of photometric redshifts for weak lensing surveys} 
\citep{Gatti22,Cawthon22,Rau23},  {relying on the fact that sources with an unknown redshift distribution (i.e. the GW sources) will cluster more strongly when cross-correlated with galaxies at the peak of the unknown sources' redshift distribution. }

By using the clustering redshift of the GW sources \citep{PhysRevD.93.083511,Mukherjee:2018ebj, Mukherjee:2019wcg, Mukherjee:2020hyn,Bera:2020jhx}, we can measure the cosmic expansion history after marginalizing over the GW bias parameters  {encoding the redshift evolution of the bias}. Apart from cross-correlation techniques, statistical host identification \citep{Schutz,Soares-Santos:2019irc, Abbott:2020khf} and GW mass distribution \citep{Taylor:2011fs,Farr:2019twy, Mastrogiovanni:2021wsd} can also be used to infer redshifts for BBHs. However, the mass distribution of the BBHs can have intrinsic redshift dependence that influences parameter estimation, if the full mass distribution gets affected \citep{Mukherjee:2021rtw,Ezquiaga:2022zkx}. 

LVK dark standard sirens have been used to measure
the expansion history using O1+O2 data \citep{Abbott:2019yzh}  and O1+O2+O3 data \citep{LIGOScientific:2021aug}, in tandem with  \texttt{GLADE} \citep{Dalya:2018cnd} and \texttt{GLADE+} \citep{Dalya:2021ewn} for statistical host identification for a fixed cosmological population. The latest LVK measurement yields $H_0= 68_{-6}^{+8}$ km/s/Mpc (68.3\% highest density interval (HDI)) after combining with the bright siren GW170817  \citep{LIGOScientific:2021aug}. Other independent measurements of $H_0$ using statistical host identification have also been performed on the GW data \citep{Finke:2021aom,Palmese:2021mjm}.

 {In this paper, we apply the cross-correlation technique on the GWTC-3 catalog of the LVK collaboration}  \citep{LIGOScientific:2021djp} and the photometric galaxy surveys 2MPZ \citep{Bilicki:2013sza} and WISE-SuperCOSMOS (WSC) \citep{Bilicki:2016irk}  {and attempt to infer the Hubble constant $H_0$ assuming flat Lambda Cold Dark Matter (LCDM) as the baseline model. Though currently, we cannot detect clustering between GW sources and galaxies due to the limited number of GW sources and large sky localization error, this is the first application of this technique on data. The current measurement is limited by the lack of high redshift galaxies and the small number of well-localized GW sources. However, the cross-correlation technique does not depend on assumptions about the position of the lower limit of the pair-instability supernovae (PISN) mass gap to measure the Hubble constant and provides an independent technique to infer the value of the Hubble constant after marginalizing over assumptions for the GW source population. Further, we note that the cross-correlation technique does not depend on assumptions about the completeness of the underlying galaxy survey, as in the statistical host identification method.  Our work therefore complements existing $H_0$ constraints from GWTC-3 dark sirens, as it makes different assumptions and has different sources of systematic error.}


\section{Method} \label{sec:method}

 {GW sources and galaxies are tracers of the underlying dark matter distribution, with their power spectra related to the matter power spectrum by a linear bias on large scales. The angular cross-correlation between the GW sources and galaxies with known redshifts is therefore proportional to the linear bias of the GW sources, the (known) linear galaxy bias and redshift distribution, and, crucially, the unknown GW redshift distribution} \citep{Newman:2008mb, Menard:2013aaa, Schmidt:2013sba}.  {In this way, by measuring the cross-correlation of the GW sources and galaxies, we can infer the redshift distribution of the GW sources, as shown in} \cite{Mukherjee:2018ebj,Mukherjee:2020hyn,Mukherjee:2020mha}.


 {From the observed spatial distribution of GW sources (or galaxies) $n_{X}(\theta, \phi)$, we can construct a density map as 
\begin{equation}\label{delta_gw}
    \delta^{X} (\theta, \phi)=  \frac{n_{X}(\theta, \phi)}{\bar n_{X}} -1,
\end{equation}
where $\bar n_{X}$ is the mean density and $X\in\{\text{GW, g}\}$. 
The angular correlation between a galaxy map and a GW map in the spherical harmonic basis can be written as 
\begin{equation}\label{eqcl}
    \hat C^{XY}_\ell = \frac{\sum_{m}\tilde \delta^{X}_{\ell m}\tilde \delta^{*Y}_{\ell m}}{2l+1},
\end{equation}
where $\hat C^{XY}_\ell$ denotes the pseudo auto (for $X=Y$) and cross (for $X\neq Y$) angular power spectrum obtained from the masked density maps, denoted by $\tilde \delta$. One can construct the binned average power spectrum as 
\begin{equation}
\hat C^{XY}_{\ell_b}= \sum _{\ell\in l_b} W_\ell C^{XY}_\ell, 
\end{equation}
where $W_\ell$ denotes the normalised window function. The corresponding covariance matrix $\mathbf{\Sigma}$ for the angular power spectrum in the Gaussian limit can be written as 
\begin{equation}\label{eqcl2}
    \mathbf\Sigma_{\ell \ell'} = \delta_{\ell\ell'} \frac{(C^{XX}_{\ell} + n^X_{\ell})(C^{YY}_{\ell}++ n^Y_{\ell})+ C^{XY}_{\ell}}{(2\ell+1)f_{\rm{sky}}\Delta \ell},
\end{equation}
where $n^{X}_\ell$ denotes the shot noise for tracer $X$, equal to the inverse of its number density. For the cross-power spectrum, the shot noise is zero. $f_{\rm sky}\equiv \Omega_s/4\pi$ denotes the overlapping sky fraction between GW sources and galaxy catalog, and $\Delta \ell$ denotes the bin width in $\ell$-space over which we estimate the band-averaged power spectrum. Choosing a large bin width makes it possible to reduce the correlation between different multipoles arising from the mask.}  

We estimate the cosmological parameters, Hubble constant $H_0$ and matter density $\Omega_m$, along with the GW bias parameter $b_{GW}(z)= b_{GW}(1+z)^\alpha$ using a Bayesian framework based on previous works \citep{Mukherjee:2020hyn,Mukherjee:2020mha,2022MNRAS.511.2782C}.  {The posterior on the parameters given the GW data $\vec{\vartheta}_{GW} \in \{ d^i_l, \Delta \Omega^i\}$ composed of the luminosity distance $d^i_l$ and sky map $\Delta \Omega^i$ of the GW sources (denoted by the index $i$). These sources are then distributed in N$_{GW}$ bins of the GW luminosity distance. The galaxy data for $N_{\rm{gal}}$ galaxies is $\vec d_{g} \in \{ z^j, \rm{RA}^j, \rm{Dec}^j\}$ composed of the galaxy redshift and the sky position denoted by Right Ascension (RA) and declination (Dec), where $j \in \{1, N_{\rm{gal}}\}$.  After marginalizing over the nuisance parameters $\Theta_n \in \{b_{GW}, \alpha\}$, we can write the posterior on the cosmological parameters (denoted by $\Theta_c$) as} 
\begin{align}\label{posterior-1}
   \mathcal{P}(\Theta_c|\vec{\vartheta}_{GW}, \vec d_{g})&\propto  \iint d\Theta_n\, dz \, \prod_{i=1}^{N_{GW}} \Pi(z) \Pi(\Theta_n)\Pi(\Theta_c) \nonumber \\ & \times \mathcal{L}(\vec{\vartheta}_{GW}| \{C_\ell^{gg}(z)\}, \Theta_n, \vec d_g(z))\nonumber \\ & \times \mathcal{P}(\vec d_g(z)| \{C_\ell^{gg}(z)\}) \nonumber\\ & \times
   \mathcal{P}({\{d^i_\ell\}}_{GW}|z, \Theta_c, \{\hat \theta^i,\, \hat \phi^i\}_{GW}),
\end{align}
where the likelihood $\mathcal{L}(\vec{\vartheta}_{GW}| \{C_\ell^{gg}(z)\}, \Theta_n, \vec d_g(z))$ is written as 
\begin{align}\label{likeli-1}
  \mathcal{L}(\vec{\vartheta}_{GW}| &\{C_\ell^{gg}(z)\}, \Theta_n, \vec d_g(z))  =\frac{1}{\sqrt{2\pi{\mathbf\Sigma^{-1}_{{\ell}_b{\ell'}_b}}}}  \\ \nonumber & \times \exp\bigg(- 0.5\sum^{\ell_{\rm max}}_{{\ell}_b, {\ell'}_b}{{D}({\ell}_b, z){\mathbf\Sigma^{-1}_{{\ell}_b{\ell'}_b}}{D}(\ell'_b,z)}\bigg),
\end{align}
 {here, $\ell_{\rm max}$ denotes the maximum value of the multipoles that can be used (which depends on the sky localization error) and  
\begin{equation}
{D}(\ell,z) = \hat C_\ell^{GW\,g}(z)- C_\ell^{GW\,g}(z),
\end{equation}
where $ C_\ell^{GW\,g}$ is the theory model for the GW-galaxy cross-correlation and $\hat C_\ell^{GW\,g}$ is the measured cross-correlation. In this analysis, we have only considered the diagonal covariance matrix, leading to simplification in Eq.~\eqref{likeli-1}. However, in the future with more GW sources, additionally including the small off-diagonal elements of the full covariance matrix will be appropriate.}

The angular cross-correlation power spectrum $\hat C_\ell^{GW\,g}(z)$ is
obtained from cross-correlating GW sources detected above a network-matched filtering SNR with galaxy catalogs $\mathbf{d}_g(z)$. The theoretical angular cross-power spectrum is written in terms of the measured galaxy auto-power spectrum $C_\ell^{gg}(z)$, the galaxy bias $b_{g}(z$) and GW bias  $b_{GW}(z)$ as 
\begin{equation}
   C_\ell^{GW\,g}(z)= \frac{b_{GW}(z)}{b_{g}(z)} C_\ell^{gg}(z).  
\end{equation}
We describe in Appendix~\ref{gal-auto} in detail the procedure followed to measure galaxy auto-correlation and galaxy bias. The term $\mathcal{P}(\vec d_g(z)| \{C_\ell^{gg}(z)\})$ denotes the galaxy density field given the auto-power spectrum $C_\ell^{gg}(z)$. The likelihood on the luminosity distance given the cosmological parameters $\Theta_c$ and redshift is denoted by $\mathcal{P}({\{d^i_\ell\}}_{GW}|z, \Theta_c, \{\hat \theta^i,\, \hat \phi^i\}_{GW})$, and the prior on the redshift, cosmological parameters, and nuisance parameters are denoted by $\Pi(z)$, $\Pi(\Theta_c)$, and $\Pi(\Theta_n)$ respectively. 

We include only large scales $\ell\leq30$ and consider different choices of bins $\Delta \ell= 5$ and $15$ in the analysis. 
The smallest scale used $\ell_{\rm max}=30$ is chosen due to the poor sky localization of the GW sources, which imposes a beam that effectively smears out smaller scales.
We do not use the first $\ell$-bin in the analysis to minimize low-$\ell$ contamination.
Our results will also depend on $z_{\rm max}$, the maximum redshift of the flat prior on GW source redshift, $\Pi(z)$, i.e. the maximum redshift up to which Eq.~\eqref{posterior-1} is integrated.
We also choose two different values of $z_{\rm max}$, namely $z_{\rm max}=0.5$ and $z_{\rm max}= 2$ in the analysis.

\begin{figure}
    \centering
    \includegraphics[width=1.\linewidth]{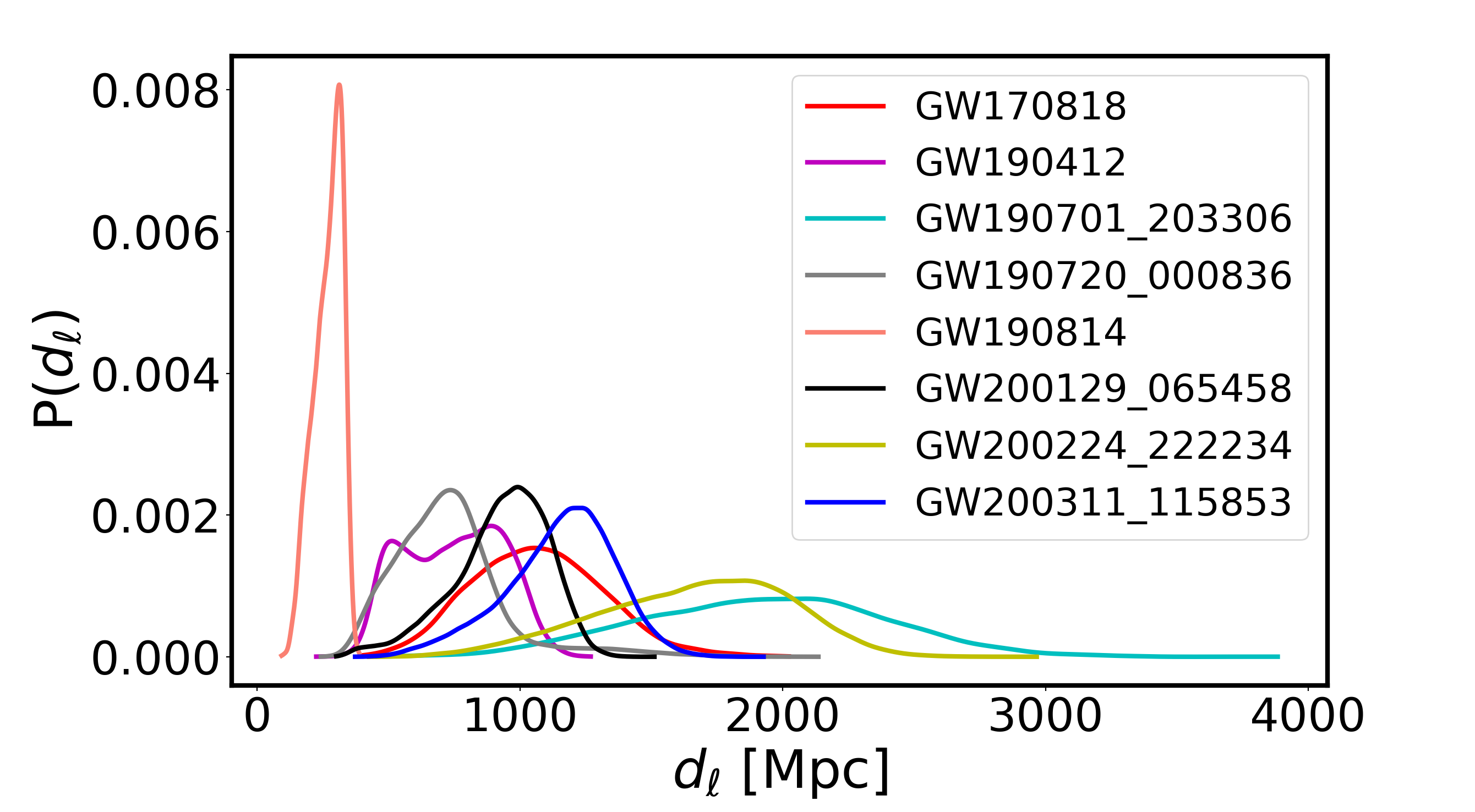}
    \caption{The luminosity distance of the eight selected GW sources from GWTC-3.}
    \label{fig:gwtc-dl}
\end{figure}
\section{GW catalog and galaxy catalog selection function} \label{sec:selection}
In this analysis, we use the publicly available GW catalog GWTC-3 detected by the LVK collaboration \citep{LIGOScientific:2021djp}. As the most constraining estimations of cosmological parameters can be made from sources having a high matched-filtering SNR, we select samples from GWTC-3 with SNR $\geq$ 11. Also, as the cross-correlation technique is most effective for sources with better sky localization error, we further select sources with sky localization error $\Delta \Omega \leq 30$ sq. deg at $68.3\%$ CI. These two selections lead to a total of eight GW events, namely GW170818, GW190412, GW190814, GW190701$\_$203306, GW190720$\_$000836, GW200129$\_$0065458, GW200224$\_$222234, and  GW200311$\_$115853. 

\begin{figure}
    \centering
    \includegraphics[width=\linewidth]{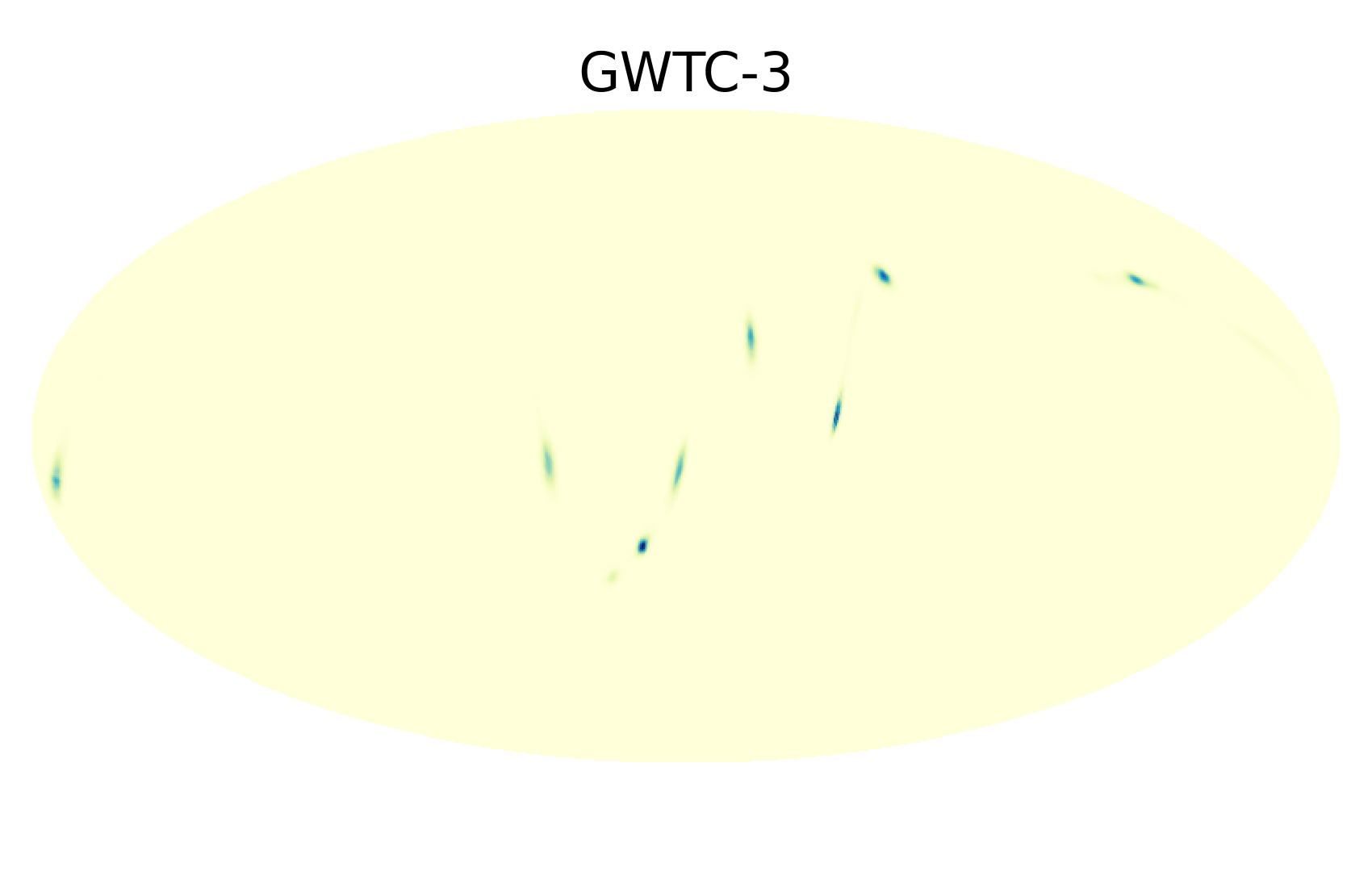}
    \includegraphics[width=\linewidth]{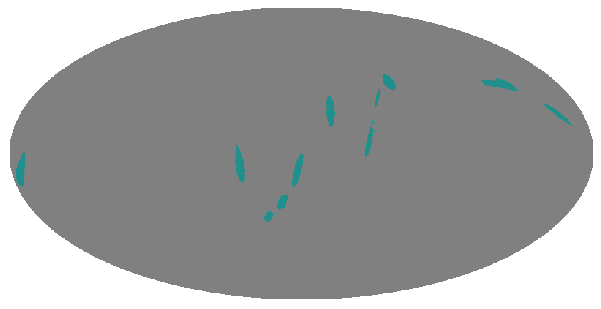}
    \caption{The sky map in equatorial coordinates of the eight selected GW sources from GWTC-3 (top) and the GW sky mask (bottom) used in the analysis.}
    \label{fig:gwtc-3}
    \end{figure}

The posteriors on the luminosity distance are shown in Fig.~\ref{fig:gwtc-dl} and the sky map of the GW sources and the sky mask are shown in Fig.~\ref{fig:gwtc-3}.  {The spatial distribution of the GW sources in the sky with good sky localization error depends on the number of GW detectors operational at the time of observation, the noise and antenna pattern of the individual detectors, and also on the properties of the individual GW sources. For sources with higher masses, the sky localization error is improved.} Also, for sources with unequal masses, the sky localization error will be better. In our analysis, we consider only the fraction of the sky ($f_{\textrm{sky}} \sim 2\%$) that's allowed by the well-localized sources for which the sky localization error is less than 30 sq. deg for O3 sensitivity of GW detectors. 

We construct three GW maps from the selected GW samples composed of Set-1 (GW190814), Set-2 (GW170818, GW1901412, GW190720\_000836, GW2001129\_065458, GW200311\_115853),  and Set-3 (GW190701\_203306, GW200224\_222234). These maps are constructed based on their luminosity distance distribution. Sources with a similar maximum value of the posterior distribution are combined to enhance the cross-correlation signal.

    \begin{figure}
    \includegraphics[width=\linewidth]{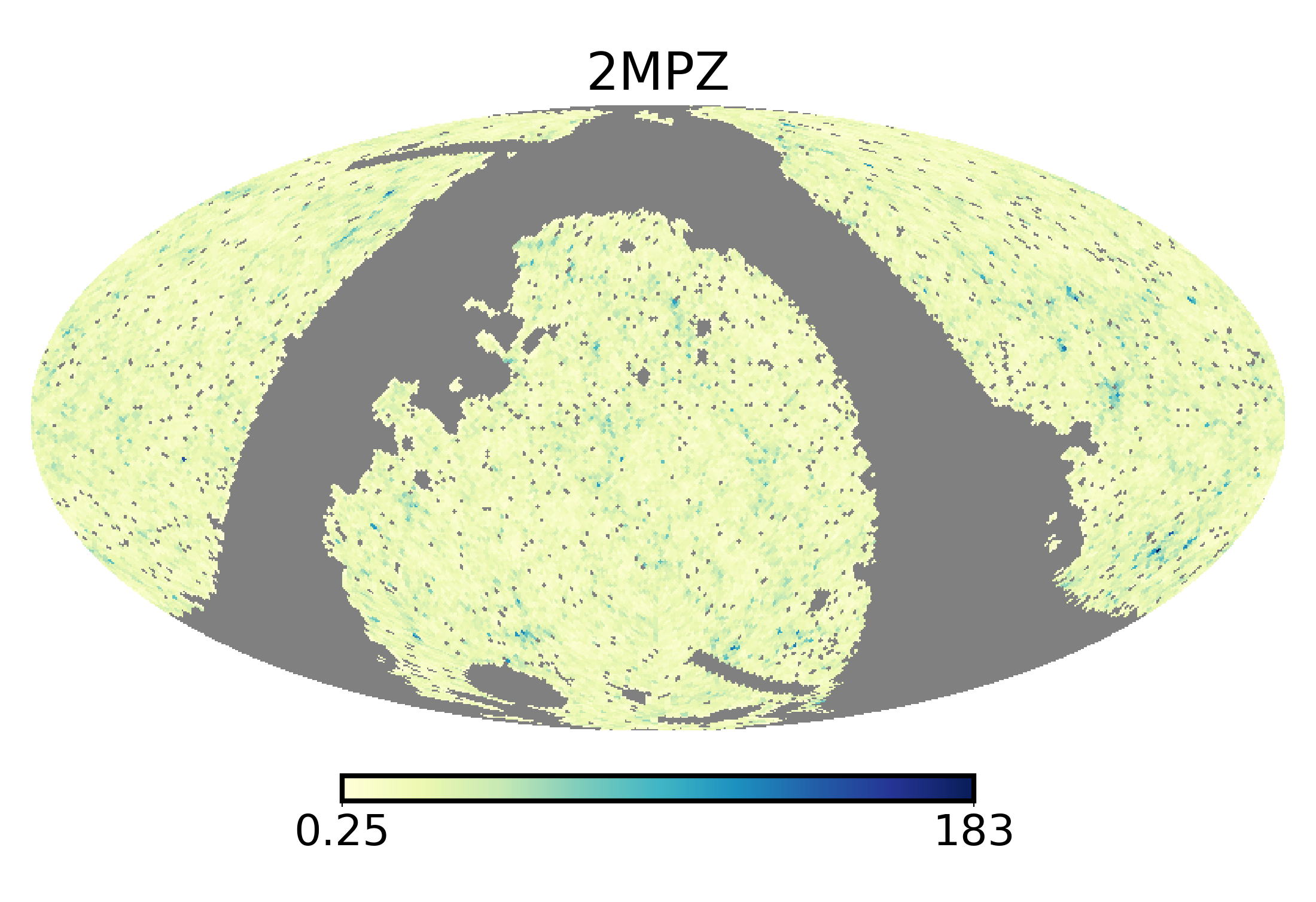}
    \includegraphics[width=\linewidth]{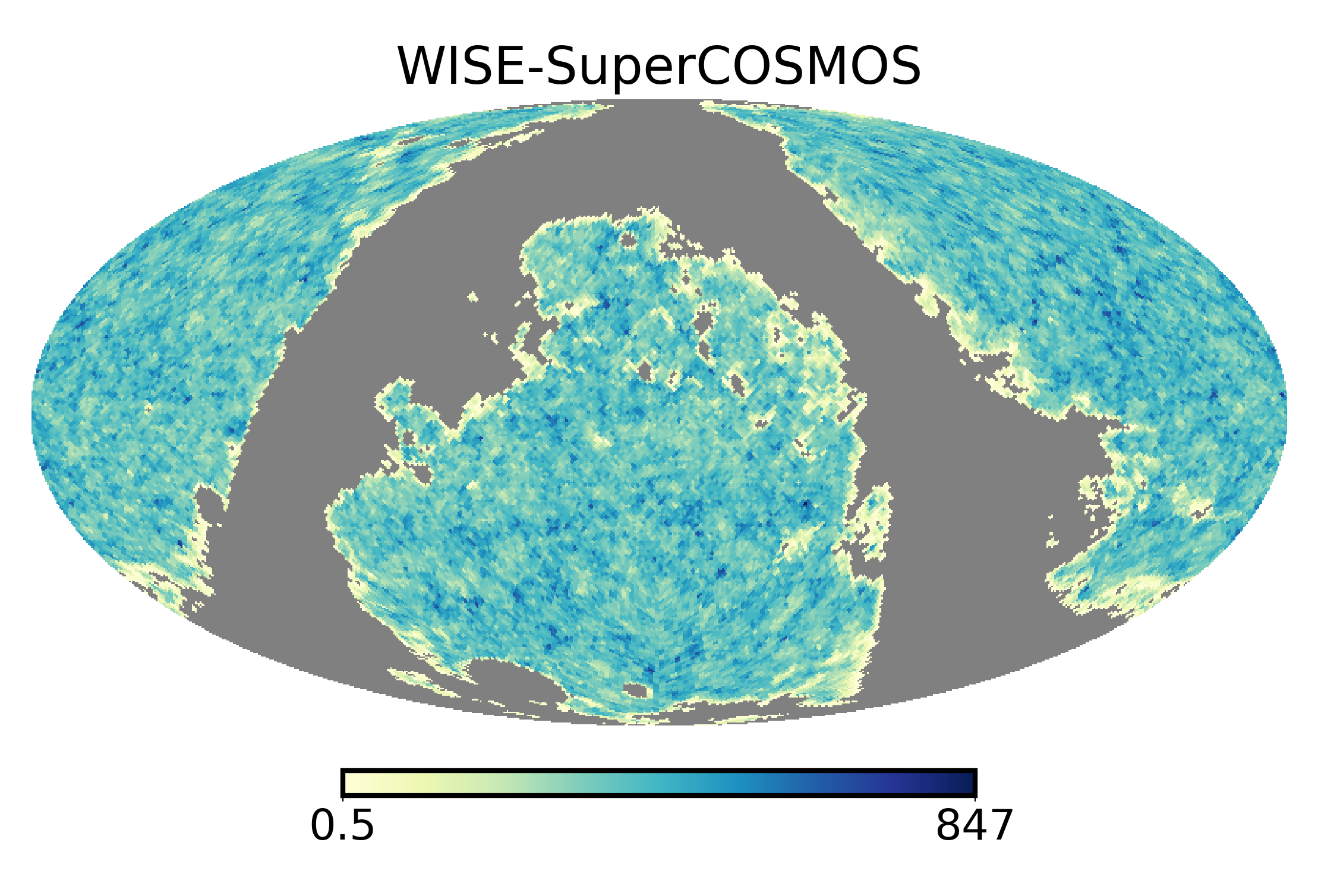}
    \caption{The sky map along with mask (gray) in equatorial coordinates of 2MPZ (top) and WSC (bottom).}
    \label{fig:2MPZ}
\end{figure}

\textit{Galaxy catalog and selection function:} 
We use galaxies
from the 2MASS Photometric Redshift catalog \citep[2MPZ][]{Bilicki:2013sza} and WISE cross SuperCOSMOS Photometric Redshift catalog \citep[WSC, ][]{Bilicki:2016irk}. The sky maps of the galaxies are shown in Fig.~\ref{fig:2MPZ}. The details of both these catalogs can be found in \cite{Bilicki:2013sza} and \cite{Bilicki:2016irk}.
The choice of mask\footnote{The details for the construction of the mask are given in Appendix~\ref{app:masks}.} for the 2MPZ \citep{Bilicki:2013sza} and WSC \citep{Bilicki:2016irk} galaxy surveys left nearly $65\%$ of the sky. 
DeCALS \citep{Zhou:2020nwq}  {is another survey that would be desirable for this analysis, as it lies at higher redshift than WSC and has considerably lower stellar contamination. However, DeCALS is focused on the Northern sky, and while the sky fraction is still high ($\sim50\%$) it
unfortunately misses
four out of the eight GW events.
As a result, we use only 2MPZ and  WSC in this paper.}   

\begin{figure}
    \centering
    \includegraphics[width=1.\linewidth]{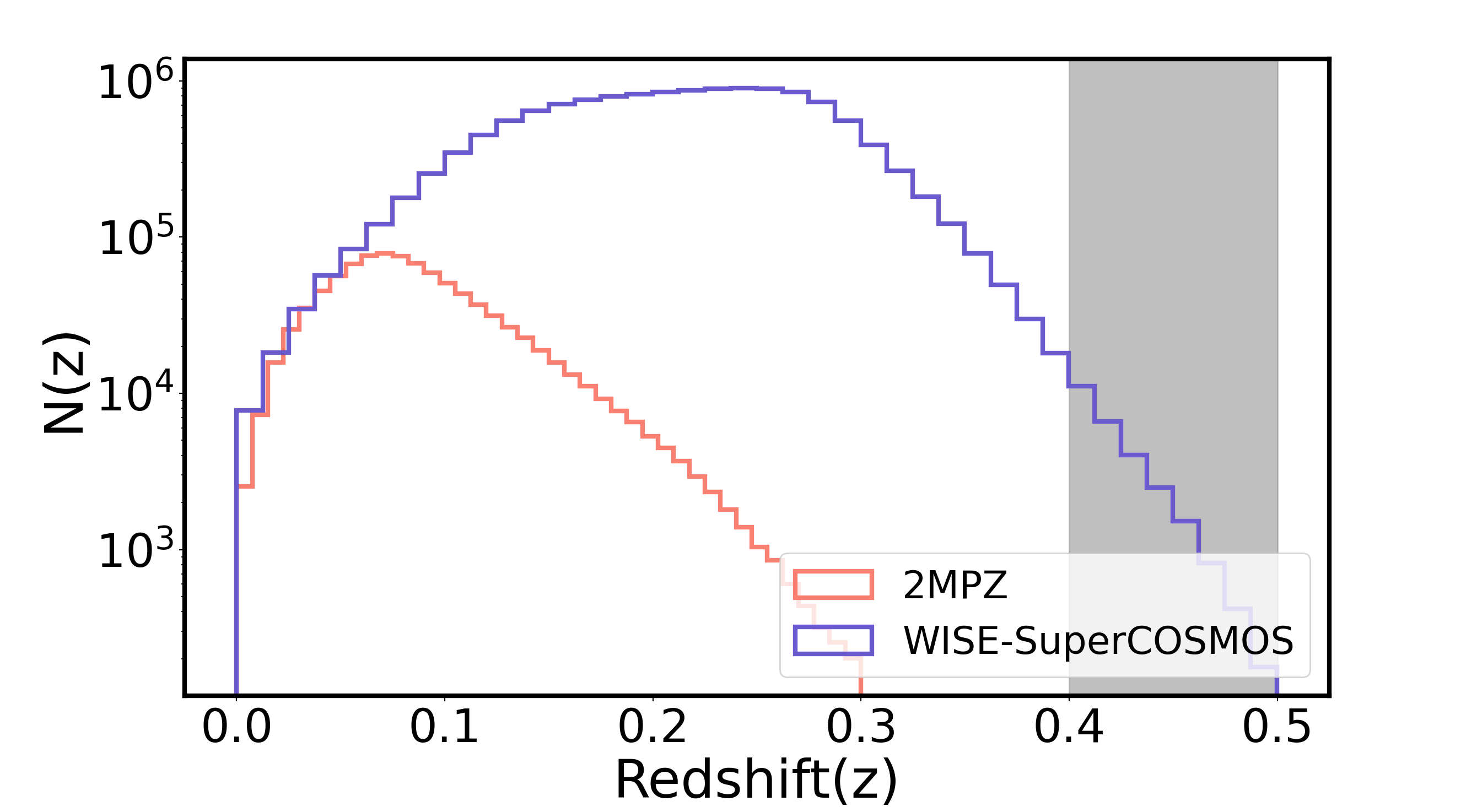}
    \caption{The redshift distribution of the 2MPZ (in orange) and WSC (in blue) galaxies. The shaded region shows the redshift range that is shot noise-dominated.}
    \label{fig:nz}
\end{figure}

The redshift distributions of 2MPZ and WSC are shown in Fig.~\ref{fig:nz} in orange and blue respectively. 
At $z < 0.1$ we use 2MPZ despite its lower number density,
as it has more precise photometric redshifts ($\sigma_z= 0.015$ \citep{Bilicki:2013sza}) and far less stellar contamination. For $z>0.1$, we use WSC exclusively (having $\sigma_z/(1+z)= 0.033$ \citep{Bilicki:2016irk}). 
At $0.3 < z < 0.4$, WSC clustering is shot noise dominated at $\ell > 60$, and is shot noise dominated at all $\ell$ for $z > 0.4$. We still measure galaxy clustering at $0.4 < z < 0.5$, albeit with increased errors (total SNR $\sim 4.7$, after subtracting shot noise, over the relevant scales $10 < \ell < 40$), but exclude $z > 0.5$ where
there are very few WSC galaxies. For both surveys,
photometric redshifts are trained using the ANNz algorithm \citep{Collister04}, 
yielding typical redshift errors $\sigma_z = 0.015$ for 2MPZ
and $\sigma_z/(1+z)= 0.033$ for WSC.

\section{Results} \label{sec:results}
We have adopted the following uniform prior ranges: $\Pi(H_0)= \mathcal{U}[20, 120]$ km/s/Mpc, $\Pi(\Omega_m)= \mathcal{U}[0.1, 0.4]$, $\Pi(z)= \mathcal{U}[0, z_{\rm max}]$, $\Pi(b_{GW})= \mathcal{U}[0.1, 6]$, and $\Pi(\alpha)= \mathcal{U}[-2, 2]$  {and Flat LCDM cosmological model}. In this analysis, we have used the \texttt{emcee: The MCMC Hammer}  \citep{2013PASP..125..306F} for estimating the posteriors of the parameter with the nwalker=40 and chain size $10^4$.We have shown results for two different choices of redshift bin-width $\Delta z=0.05$ and $\Delta z=0.1$, which are roughly 1.5 and 3 times the WSC photo-$z$ error respectively. We also consider two different choices of bins
$\Delta \ell= 5$ and $15$,
and two different values of $z_{\rm max}$, namely $z_{\rm max}=0.5$ and $z_{\rm max}= 2$.
$z_{\rm max}=0.5$ is the maximum redshift of the galaxy catalog; however, the GW sources may lie at higher redshift. $z_{\rm max} = 2$ sufficiently high beyond which detection of sources with matched filtering SNR $>12$ for a prior range of $H_0$ between [20, 120] km/s/Mpc with O3 detector sensitivity will not be feasible for these eight sources with their detector frame mass parameters. 
To make sure that the result is not susceptible to the choice of the maximum value of mass (i.e. population independent), we also checked that our results do not change even when we take $z_{\rm max}=4$.  A null test validating the technique on a mock random catalog (without any spatial clustering) is shown in Appendix~\ref{validation} with the same sky mask of the GW data and galaxy catalog. This shows that our method is unbiased and exhibits no constraints from a random galaxy catalog. 

\begin{figure}
    \centering
\includegraphics[width=1.\linewidth]{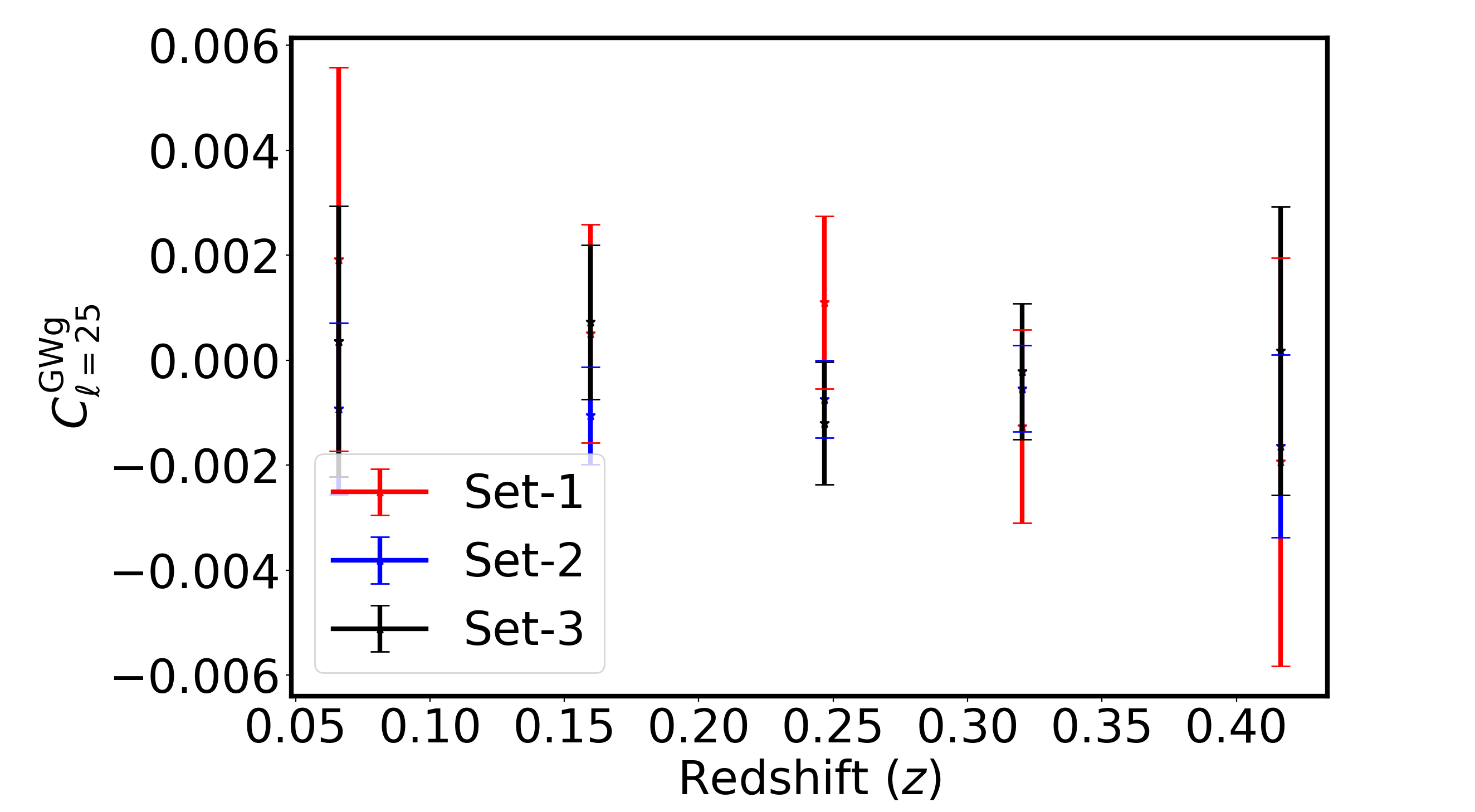}
    \caption{ {The band-averaged cross-correlation power spectrum between GW sources and galaxies with a $\Delta \ell=15$ bin-width and $\Delta z=0.1$ is shown for three different maps of the GW sources Set-1 (GW190814), Set-2 (GW170818, GW1901412, GW190720\_000836, GW2001129\_065458, GW200311\_115853),  and Set-3 (GW190701\_203306, GW200224\_222234) composed from the selected eight events, as a function of the median value of the redshift bin.}}
    \label{fig:cls}
\end{figure}

 {The cross-correlation between the GW sources and galaxies is dominated by shot noise and we do not measure any cross-correlation clustering signal with statistical significance. We show the measured band average cross-correlation signal $\hat C_\ell^{GW g}$ as a function of redshift in} Fig.~\ref{fig:cls}  {for $\Delta \ell=15$ and $\Delta z=0.1$ with the three GW maps constructed from the eight events.   The diagonal error bars are shown on the measured values. The band average signal shows a signal that is consistent with zero at all the redshift bins.} 
\begin{table}
\large
\centering
\begin{tabular}{|c| c| c| c| c|}
 \hline
 \multirow{2}{*}{$\Delta l$} & \multirow{2}{*}{$\Delta z$} & \multirow{2}{*}{$z_{\rm max}$} &  $H_0 (\rm{dark})$
 & $H_0(\rm{dark + bright})$ \\
 & & & [km s$^{-1}$ Mpc$^{-1}$] & [km s$^{-1}$ Mpc$^{-1}$] \\
 \hline\hline
 5 & 0.1 & 0.5 & $71.1_{-13}^{+27}$ & $67.7_{-5}^{+18}$\\ 
 \hline
 15 & 0.1 & 0.5 & $67.5_{-6}^{+25}$ & $66.9_{-4}^{+6}$\\
 \hline
 15 & 0.1 & 2 & $79.8_{-15}^{+23}$ & $71.0_{-6}^{+11}$\\
 \hline
 15 & 0.05 & 0.5 & $78.5_{-21}^{+17}$ & $73.0_{-5}^{+11}$ \\
 \hline
 15 & 0.05 & 2 & $82.4_{-27}^{+23}$ & $75.4_{-6}^{+11}$\\[1ex] 
 \hline
\end{tabular}
 \caption{The median and 68.3\% ETI values of $H_0$ are shown for dark and dark+bright sirens for  different choices of parameters such as $C_\ell$ bin-width $\Delta l$, redshift bin-width $\Delta z$, and maximum redshift in the prior $z_{\rm max}$. The values for all the choices are consistent with each other within the error bars.}\label{tab:1}
 \end{table}

\begin{figure}
    \centering
    \includegraphics[trim={0.3cm 0.cm 0.8cm 0.2cm},clip,width=1.08\linewidth]{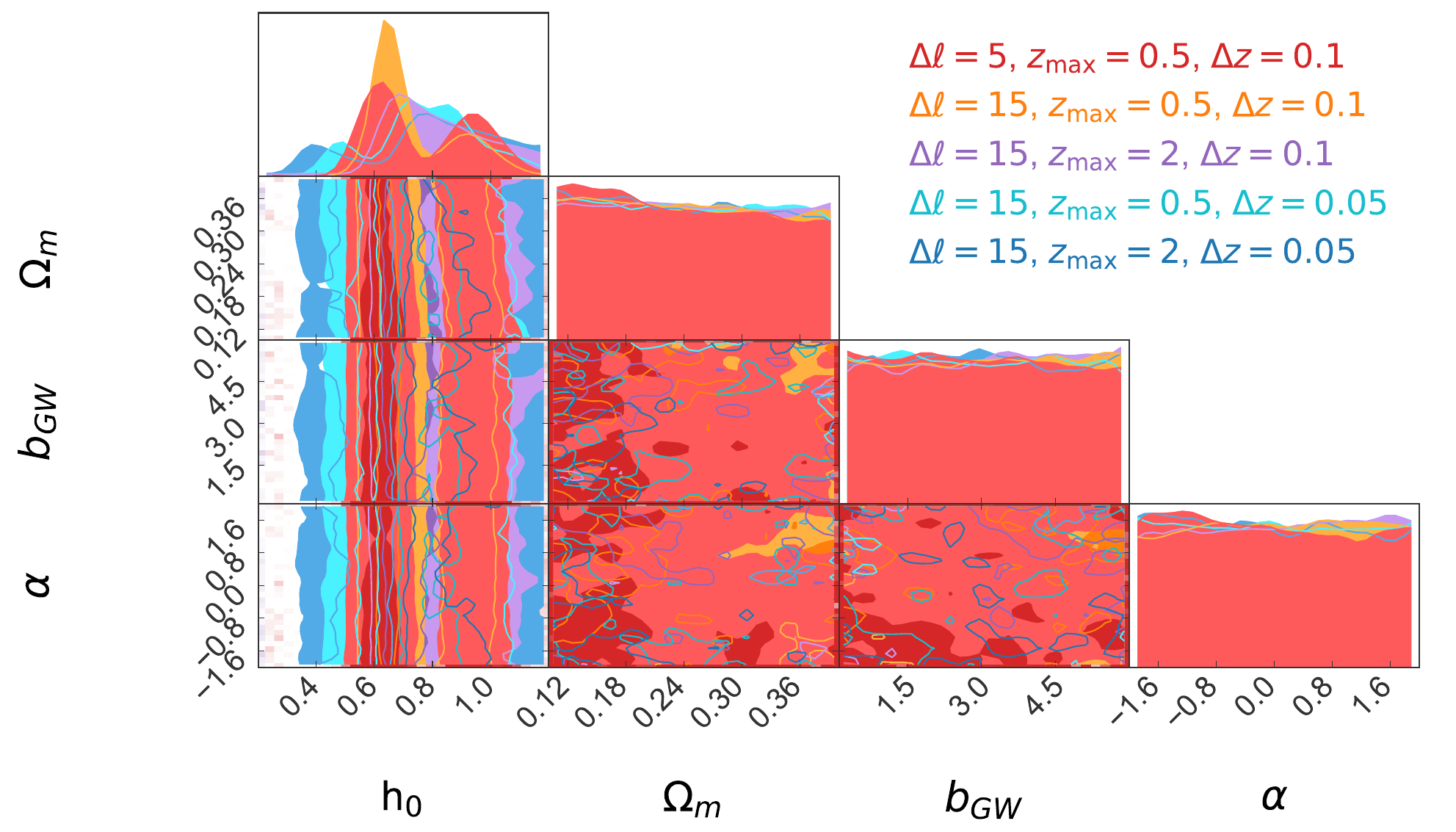}
     \caption{ {The joint constraints on $H_0= 100h_0$ km/s/Mpc, $\Omega_m$, and $b_{GW}(z)= b_{GW}(1+z)^\alpha$ for different choices of bin-width $\Delta l$, $z_{\rm max}$, and $\Delta z$.}}
    \label{fig:h0}
\end{figure}

 {The joint estimation of the Hubble constant along with the matter density and GW bias parameters are shown in Fig.~\ref{fig:h0}. Constraints on the Hubble constant are bimodal, with the median value $H_0= 82.4_{-27}^{+23}$ km/s/Mpc (the upper and the lower limit indicates the $68.3\%$ equal-tailed interval (ETI)) for $\Delta z=0.05$ and the maximum redshift in prior $z_{\rm max}=2$. The constraints for different choices of $\Delta l$, $\Delta z$, and $z_{\rm max}$ are shown in Table \ref{tab:1}. Due to the limited number of GW sources, we are not able to detect the cross-correlation signal with galaxies, and hence no statistically significant inference of $H_0$ is possible currently. The value of the matter density $\Omega_m$ and bias parameter $b_{GW}(z)$ are unconstrained as well.}

 {The posterior on the Hubble constant from only dark sirens is uninformative for all the cases and spans nearly the complete prior range due to the non-detection of the cross-correlation signal. In this first application of the cross-correlation technique on GW data, the measurements of $H_0$ presented are from a small number of GW sources. As a result, the clustering of the GW sources is not measured with any statistical significance, resulting in a weak estimate of $H_0$. The values in Table \ref{tab:1} show that all the estimates are statistically consistent due to large error bars. 
The measurement in this work agrees with the dark siren measurement including population assumption of  $H_0=67_{-12}^{+13}$ km/s/Mpc (68.3\% HDI) by the LVK collaboration \citep{LIGOScientific:2021aug}. This measurement is also limited by the small number number of sources used \citep{LIGOScientific:2021aug} and systematics related to population assumptions.}

 {To test the robustness of our results, we have checked the following aspects, (i) randomly varied the galaxy bias parameter within their error bars, (ii) enhanced the covariance matrix by a factor of four, (iii) changed the cosmological parameters such as $S_8\equiv \sigma_8 \sqrt{\Omega_m}$ that is used to fit the galaxy power spectrum $C_\ell^{gg}$ from $S_8=0.832$ (Planck-2018 \citep{Aghanim:2018eyx}) to a lower value $S_8=0.75$ as indicated by the KiDS Collaboration  \citep{2021A&A...646A.140H}, (iv) changed the value of $H_0=67$ km/s/Mpc to $H_0=74$ km/s/Mpc \citep{Riess:2019cxk} to estimate the galaxy bias parameters, (v) changed the galaxy sample selection by additionally
removing WSC sources with a low probability of being galaxies using the SVM catalog of \cite{Krakowski16}, requiring $p_{\rm gal} > 0.67$, (vi) changed the redshift bin width $\Delta z$ to 0.05 which is comparable to the photo-z errors, and (vii) change in the maximum redshift $z_{\rm max}$ in the prior. The posterior on $H_0$ did not show any significant variation for (i)--(v) cases. For scenarios (vi) and (vii), the $H_0$ posterior shows some variation. As shown in Table~\ref{tab:1}, the error bars increase with the decrease in $\Delta z$ and increase in $z_{\rm max}$. The change with $\Delta z$ happens because the galaxy redshift kernels begin to overlap due to photo-$z$ errors, violating our assumption that the GW cross-correlations in neighboring bins are uncorrelated. The increase in $z_{\rm max}$ includes the contribution from higher redshifts in the prior. As there is no support for galaxies beyond $z=0.5$, there is no information beyond the prior choice. As a result, allowed large prior on redshift (higher $z_{\rm max}$) enhances the error at high $H_0$ in comparison to low $H_0$.}

\begin{figure}
    \centering
    \includegraphics[width=1.\linewidth]{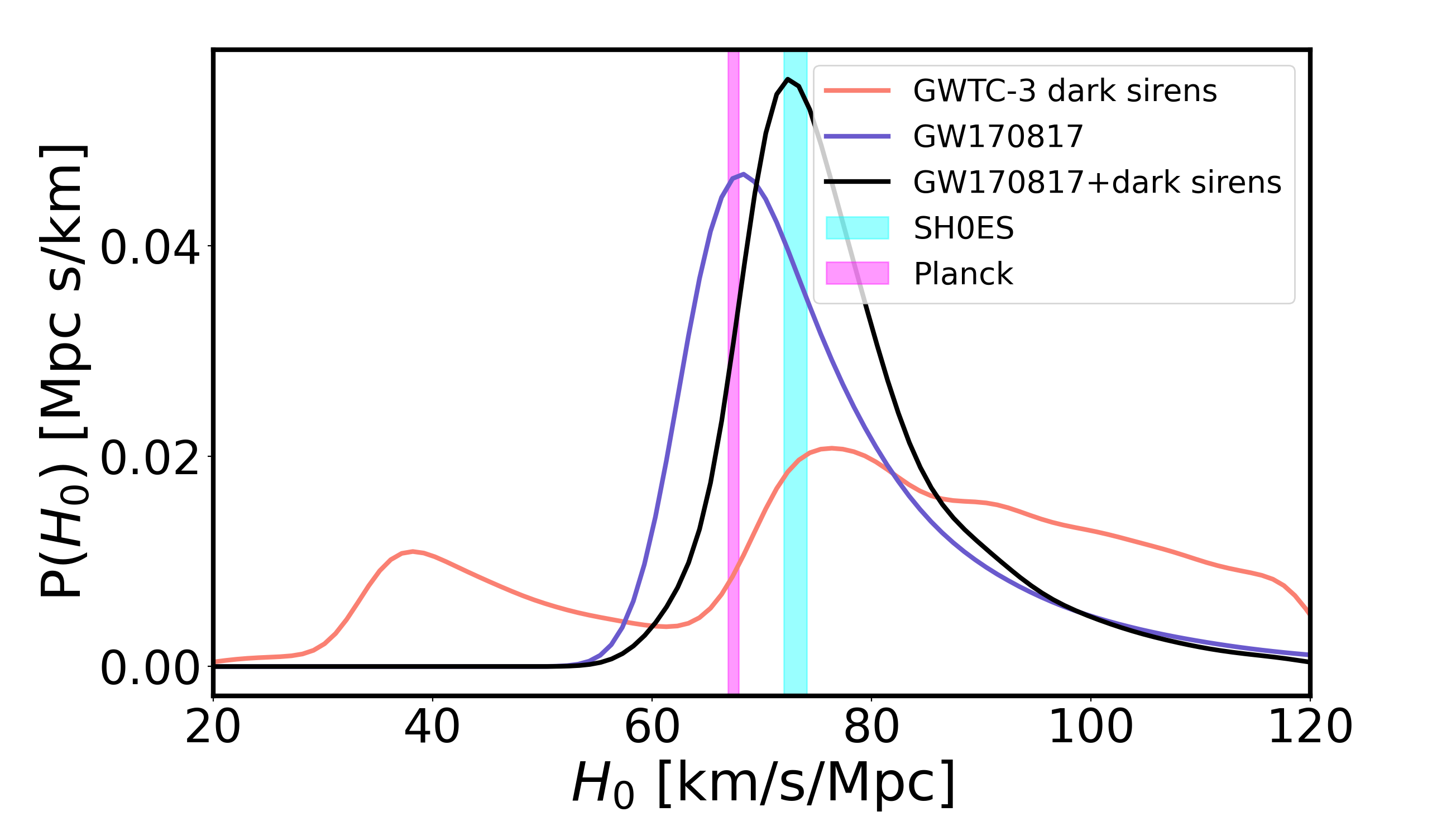}
     \caption{Hubble constant $H_0$ measurement from GWTC-3 dark sirens, bright siren GW170817, and combining the both are shown along with the mean and the standard deviation on the measurements from Planck-2018 \citep{Aghanim:2018eyx} and SH0ES \citep{Riess:2021jrx}.}
    \label{fig:h0all}
\end{figure}

 {By combining the bright standard siren measurement from GW170817 with a better measurement of peculiar velocity \citep{Mukherjee:2019qmm}, we show the corresponding posterior on $H_0$ in Fig.~\ref{fig:h0all} with the median value of $H_0= 75.4_{-6}^{+11}$ km/s/Mpc (68.3$\%$ ETI) for $\Delta z=0.05$ and $z_{\rm max}=2$. The value of $H_0$ for other choices of $\Delta z$ and $z_{\rm max}$ are mentioned in Table \ref{tab:1}. In comparison, the median and 68.3$\%$ ETI from GW170817 is $H_0= 72.8_{-8}^{+15}$ km/s/Mpc \citep{Mukherjee:2019qmm}---our combined constraint thus provides a slight improvement on the errorbar. The values of the Hubble constant are consistent with each other within about  1-$\sigma$ for all the choices of $z_{\rm max}$ and $\Delta z$. As the GW-galaxy cross-correlation is not detected, the choices of prior on redshift (such as $z_{\rm max}$ and $\Delta z$) mildly impact the results.}

In Fig.~\ref{fig:h0all} we compare the GW measurements of $H_0$ with the measurements from Planck, $H_0= 67.4^{+0.5}_{-0.5}$ km/s/Mpc \citep{Planck:2018vyg} and with the measurement of $H_0=73.04^{+1.05}_{-1.05}$ km/s/Mpc from SH0ES \citep{Riess:2021jrx}. The current measurements from the dark sirens are not sufficiently constraining yet to resolve the $H_0$ tension \citep{Verde:2019ivm, DiValentino:2021izs,Dainotti:2021pqg}. 
Though the systematic uncertainties in our measurement of $H_0$ are smaller than the statistical uncertainties, in the future with more GW sources and a better galaxy catalog, we will be able to better assess the influence of any systematic uncertainties.  

\section{Conclusion and future outlook} \label{sec:conclusions}
We present the first application of the Hubble constant $H_0$ inference from dark standard sirens using the cross-correlation technique. With the best eight sources available from GWTC-3, we obtain a median value of Hubble constant $82.4_{-27}^{+23}$ km/s/Mpc ($68.3\%$ ETI),  {with the weak constraints on $H_0$ arising from the non-detection of the GW-galaxy cross-correlation.}
In the future, with the availability of $z < 0.8$ spectroscopic galaxy catalogs such as DESI \citep{Aghamousa:2016zmz} and SPHEREx \citep{Dore:2018kgp} (supplemented by $z > 0.8$ spectroscopy from Euclid \citep{Euclid:2019clj} and photometric redshifts from Vera Rubin Observatory \citep{LSSTScienceBook}), cross-correlation of the GW sources with galaxies will be a powerful technique to measure the expansion history  \citep{Mukherjee:2018ebj,Mukherjee:2020hyn,2022MNRAS.511.2782C} and testing the general theory of relativity \citep{Mukherjee:2020mha}. The measurement presented in this analysis reports a value of the Hubble constant which is not influenced by the choice of the pair-instability supernovae (PISN) mass gap \citep{Farmer}. The dependence of GW sources on the galaxy properties through the GW bias parameter is marginalized in this analysis. In the future, with the availability of a few hundred dark sirens, the cross-correlation technique will be able to infer the clustering redshift of sources more accurately, and this will be able to shed further light on the tension in the Hubble constant determinations \citep{Verde:2019ivm, DiValentino:2021izs}.

\section{Acknowledgments}
The authors are thankful to Gergely Dalya for carefully reviewing the manuscript and providing useful comments during the LVK internal review. The authors are also thankful to Maciej Bilicki for providing insightful suggestions on the paper. Research at Perimeter Institute is supported in part by the Government of Canada through the Department of Innovation, Science and Economic Development and by the Province of Ontario through the Ministry of Colleges and Universities. SM is supported by the Simons Foundation. AK thanks the AMTD Foundation for its support. This analysis is carried out at the Symmetry computing facility of the Perimeter Institute and the Infinity cluster hosted by Institut d'Astrophysique de Paris. We thank Stephane Rouberol for smoothly running the Infinity cluster. This research has made use of data obtained from the SuperCOSMOS Science Archive, prepared and hosted by the Wide Field Astronomy Unit, Institute for Astronomy, the University of Edinburgh, which is funded by the UK Science and Technology Facilities Council. 

The authors would like to thank the  LIGO/Virgo/KAGRA scientific collaboration for providing the data. LIGO is funded by the U.S. National Science Foundation. Virgo is funded by the French Centre National de Recherche Scientifique (CNRS), the Italian Istituto Nazionale della Fisica Nucleare (INFN), and the Dutch Nikhef, with contributions by Polish and Hungarian institutes. This material is based upon work supported by NSF's LIGO Laboratory which is a major facility fully funded by the National Science Foundation.

%


\software{Astropy \citep{2013A&A...558A..33A, 2018AJ....156..123A}, emcee: The MCMC Hammer \citep{2013PASP..125..306F}, Giant-Triangle-Confusogram \citep{Bocquet2016}, healpy\citep{Gorski:2004by, Zonca2019}, IPython \citep{PER-GRA:2007}, Matplotlib \citep{Hunter:2007},  NaMaster \citep{Alonso:2018jzx}, NumPy \citep{2011CSE....13b..22V}, and SciPy \citep{scipy}.}


\appendix
\section{Measuring galaxy auto-correlation and bias}\label{gal-auto}
\subsection{Method}
From the galaxy auto-power spectrum, we infer the galaxy bias $b_g(z)$ by fitting a simple linear bias times the nonlinear ``Halofit'' matter power spectrum model \citep{Takahashi12}.
We use the NaMaster code \citep{Hivon02,Alonso:2018jzx} to measure
pseudo-$C_{\ell}$ for each redshift slice, applying a $1^{\circ}$ apodization \citep[``$C^1$ apodization'';][]{Grain09} to the galaxy mask.
For WSC, we additionally deproject
the Schlegel-Finkbeiner-Davis \citep{sfd98} dust extinction map and a stellar density map from Gaia \cite{GaiaDR1} to reduce the impact of contamination, following \cite{Koukoufilippas}.
We fit the one-parameter bias model to the data in the range $10 < \ell < 40$, with shot noise fixed.
For WSC, we additionally allow
for systematic variations in the number density
from variations in the zero point
between SuperCOSMOS plates. We add a template to the model 
\begin{equation}
   C_{\ell}^{\rm plate} = A \exp{[-2 (\ell \theta_{\rm plate})^2/12]} 
\end{equation}
where $\theta_{\rm plate}$ is the plate scale, 5$^{\circ}$ \citep{Koukoufilippas}.
Finally, we fix the shot noise to the inverse of the angular number density (in steradians) except for the $0.4 < z < 0.5$ bin, where we adjust it downwards by 5\% to match the high-$\ell$ power of $C_{\ell}^{gg}$. For the other bins, we check that $1/\bar{n}_g$ matches the high-$\ell$ power in $C_{\ell}^{gg}$, and the discrepancies are small compared to the clustering amplitude at $10 < \ell < 40$.

To model the redshift distribution when determining $b_g$, we convolve the observed photometric redshift distribution with a
Gaussian for 2MPZ \citep{Balaguera-Antolinez}
and a 
generalized Lorentzian for WSC, 
\begin{equation}
 P(\delta z) \propto \left(1 + \frac{\delta z ^2}{2 a s^2}\right)^{-a}   
\end{equation}
\citep{PeacockBilicki18}. 
The width evolves as a function of redshift, for the Gaussian following 
\begin{equation}
    \sigma = 0.027 \tanh{(-20.78 z_p^2 +7.76 z_p + 0.05)}/(1+z_p)
\end{equation}
i.e.\ increasing from 0.0013 at $z_p = 0$ to 0.013 at $z_p = 0.1$; and for the Lorentzian, 
\begin{equation}
    a(z_c) = -4z_c + 3
\end{equation}
 and 
 \begin{equation}
    s(z_c) = 0.04 z_c + 0.02 
 \end{equation}
 where $z_c$ is the midpoint of each redshift bin. Other choices for the redshift error (e.g.\ redshift-independent modified Lorentzian for 2MPZ in \citep{Bilicki:2013sza} and \citep{PeacockBilicki18}) yield very similar results.

\begin{figure*}
    \centering
    \includegraphics[width=1.\linewidth]{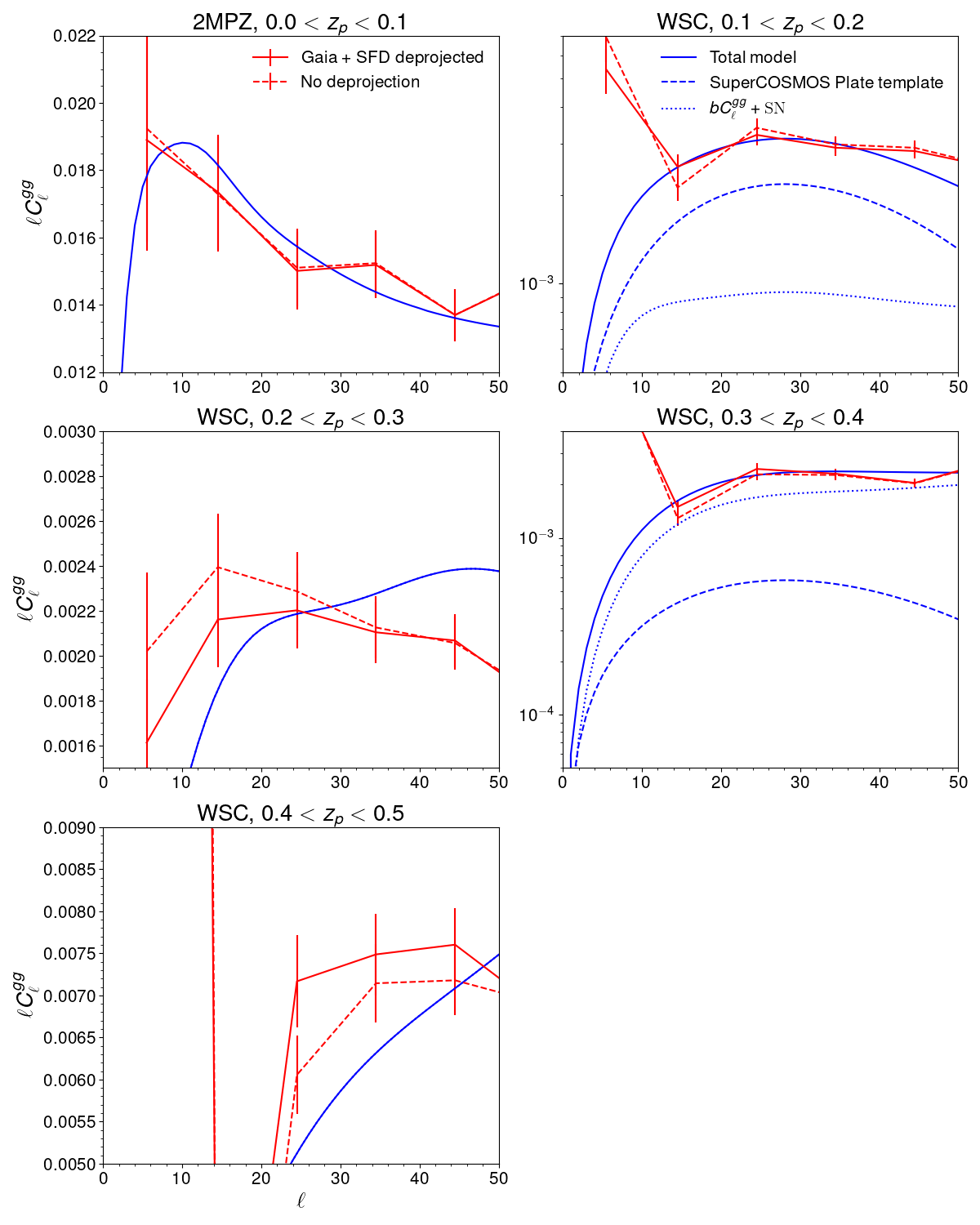}
    \caption{Galaxy auto-correlations in different photometric redshift bins, with the default measurement (deprojecting Gaia stellar template and Schlegel-Finkbeiner-Davis dust map) in solid red, and measurement with no deprojection in dashed red. The model is shown in blue: the total model is in solid blue, which for WSC is the sum of the SuperCOSMOS plate template following \citep{Koukoufilippas} (dashed blue) and $b C_{\ell}^{gg}$ plus shot noise (dotted blue).}
    \label{fig:clgg}
\end{figure*}

We assume that the galaxy bias is redshift-independent in each bin, and obtain best-fit values of $b_g(z_c = 0.15) = 0.66$, $b_g(z_c = 0.25) = 1.35$, $b_g(z_c = 0.35) = 1.76$,
and $b_g(z_c = 0.45) = 2.33$. The very low value of $b_g$ in the second bin is driven by the SuperCOSMOS plate template, which is degenerate with the cosmological contribution due to the limited multipole range considered ($10 < \ell < 40$).
{Fig.~\ref{fig:clgg} shows plots of $C_{\ell}^{gg}$for the four redshift bins of WSC. }


\subsection{Results}

We compare our bias measurement to previous results in Table~\ref{tab:2mpz_bias} for 2MPZ. For 2MPZ, the precise sample selection
is sightly different between analyses; \cite{Stolzner} use $0 < z_{\rm phot} < 0.105$, \cite{Balaguera-Antolinez} use $0 < z_{\rm phot} < 0.08$, \cite{PeacockBilicki18} use $0.05 < z_{\rm phot} < 0.1$, and \cite{Alonso15} use $0.03 < z_{\rm phot} < 0.08$ and additionally apply a bright cut of $K_s > 12$.  However, despite these subtle differences, the previous results are all broadly consistent with each other and with our result (although the ``fixed cosmology'' result from \cite{Stolzner} is significantly low compared to the rest). The differences in errors come mainly from different scales fit, except for \cite{Balaguera-Antolinez}, which have much larger errors because they allow cosmological parameters to vary without including external data (unlike \cite{Stolzner}, who include the Planck likelihood when varying cosmological parameters).

\begin{table}
    {\begin{tabular}{c|c}
    Analysis & Bias \\
    \hline
    \cite{Stolzner} fix.\ cosmo.\ & $1.03 \pm 0.03$ \\
    \cite{Stolzner} marg.\ cosmo.\ & $1.19 \pm 0.028$ \\
    \cite{Balaguera-Antolinez} & $1.14\pm 0.38$ \\
    \cite{PeacockBilicki18} & $1.18 \pm 0.009$ \\
    \cite{Alonso15} & $1.18 \pm 0.03$ \\
    \hline
    Default & $1.18 \pm 0.033$ \\
    Multiply by WSC mask & $1.18 \pm 0.017$ \\
    Mask thres.\ 0.8 & $1.18 \pm 0.016$ \\
    Mask thres.\ 0.9 & $1.22 \pm 0.018$ \\
    Lorentzian $dN/dz$ $1.21 \pm 0.032$
    \end{tabular}}
    \caption{Summary of bias results for 2MPZ with $0 < z_{\rm phot} < 0.1$. Results from the literature are on the top, and our results are on the bottom (below the horizontal line), with several variations in the mask or assumed form of $dN/dz$.}
    \label{tab:2mpz_bias}
\end{table}

{We compare our bias measurement to previous results in Table~\ref{tab:wsc_bias} for WSC.
We also test the robustness of our measurements by displaying several different variations in constructing the mask or redshift distribution.}
{For WSC, we also compare to \cite{Stolzner}, although their bins are slightly different than ours ($0.105 < z_{\rm p} < 0.21$
and $0.21 < z_{\rm p} < 0.3$).
We also compare to results from \cite{Xavier19},
who use bins of $\Delta z = 0.05$ from 0.15 to 0.35.
Therefore, their central redshift differs from ours in the first and third bins (0.175 vs.\ 0.15 and 0.325 vs.\ 0.35); in the second bin, we average their two bins in $0.2 < z_{\rm p} < 0.3$.
Similarly, we also average the $\Delta z = 0.05$ bins of \cite{PeacockBilicki18}, and report their result from the $0.3 < z_{\rm p} < 0.35$ bin in the third row.}

{Our WSC results have much larger uncertainties than any of the previous results. This is from the combination of our restricted scale cuts ($10 < \ell < 40$) and the fact that we add a template for the imprint of SuperCOSMOS plates on the large scale power, following \cite{Koukoufilippas}. Over this restricted range in scales, the cosmological contribution is nearly degenerate with the amplitude of the plate template. However, if we consider a larger multipole range, $10 < \ell < 70$, we can break the degeneracy and achieve much tighter constraints on the bias. Our results are broadly consistent with previous results, but there are differences at the 10-20\% level. This is likely due to slightly different masks and sample selection; in particular, each work uses a different method to remove stellar contamination from WSC (which is a larger problem than for 2MPZ). We are encouraged that we see the same trend of increasing bias as \cite{PeacockBilicki18}.}


\begin{table*}
    \hspace{-30pt}
     \resizebox{1.1\textwidth}{!}
     {\begin{tabular}{c|c|c|c|c|c}
  \multirow{ 2}{*}{$z_{\rm p}$ range} & \cite{Stolzner} & \cite{Stolzner}
&
    \multirow{ 2}{*}{\cite{Xavier19}} &
    \multirow{2}{*}{\cite{PeacockBilicki18}} \\
    
     & fix cosmo. & marg. cosmo. & &\\
    \hline
   $0.1 < z_{\rm p} < 0.2$ & $ 0.88 \pm 0.03$  & $0.83 \pm 0.026$ & 1.43 & 1.106 \\
   $0.2 < z_{\rm p} < 0.3$ & $ 0.80 \pm 0.02$  & $0.99 \pm 0.034$ & 1.2 & 1.175  \\
   $0.3 < z_{\rm p} < 0.4$ & & & 1.26 & 1.548
   \\
   $0.4 < z_{\rm p} < 0.5$ & & & & & \\
\hline \hline
\multirow{2}{*}{Default} &
    \multirow{2}{*}{$10 < \ell < 70$} &
    Following \cite{Novaes} &
    Following \cite{RafieiRavandi} &
    unWISE mask &
    SVM $P > 0.9$  \\
    & & $10 < \ell < 70$ & $10 < \ell < 70$
    & $10 < \ell < 70$ & $10 < \ell < 70$ \\
    \hline
   $0.66 \pm 0.44$ & $1.10 \pm 0.051$ & $1.08 \pm 0.051$ & $1.09 \pm 0.045$ & $0.99 \pm 0.056$ & $1.06 \pm 0.045$ \\
   $1.35 \pm 0.66$ & $1.30 \pm 0.047$ & $1.41 \pm 0.043$ & $1.55 \pm 0.040$ & $1.14 \pm 0.054$ & $1.57 \pm 0.038$ \\
   $1.78 \pm 0.87$ & $1.66 \pm 0.064$ & 
   $1.78 \pm 0.06$ & $1.89 \pm 0.050$
   & $1.48 \pm 0.068$ & $1.86 \pm 0.098$ \\
   $2.33 \pm 1.10$ & $2.36 \pm 0.41$ & $3.60 \pm 1.41$
   & $4.04 \pm 0.21$ & $3.75 \pm 0.19$ & $4.59 \pm 0.28$
    \end{tabular}}
    \caption{Summary of bias results for WSC. Results from the literature are on the top, and our results are on the bottom with several variations in the sample selection or scale cut.}
    \label{tab:wsc_bias}
\end{table*}

\section{Construction of galaxy mask and galaxy catalog selection}\label{app:masks} 
\subsection{2MPZ}
2MPZ is derived from the all-sky 2MASS near-infrared extended source catalog (XSC) \citep{Jarrett00,Skrutskie06}, cross-matched to the infrared AllWISE \citep{Wright10} and optical SuperCOSMOS catalogs \citep{Hambly01a,Hambly01b,Hambly01c,Peacock16}.

The galaxy masks were carefully
constructed to remove areas with large numbers of stars or other systematics
that could affect galaxy clustering, either by direct stellar contamination
or by correlations, e.g.\ suppressed galaxy density in regions of high stellar density or extinction.
We follow \cite{Balaguera-Antolinez} to construct the 2MPZ mask, starting
by masking low Galactic latitudes ($|b| < 10^{\circ}$, areas of high galactic
extinction ($E(B-V) > 0.3$ from \cite{sfd98}, and areas of high stellar density
as estimated from the 2MASS Point Source Catalog ($\log{n_{\rm star}} > 3.5$)\footnote{\href{https://www.ipac.caltech.edu/2mass/releases/allsky/ doc/sec4\_5c.html}{https://www.ipac.caltech.edu/2mass/releases/allsky/ doc/sec4\_5c.html}}.
We further include manual cutouts around the LMC and SMC, excluding $275.47 < {\rm RA} < 285.47$ and $-37.89 < {\rm DEC} < -27.89$, and $300.81 < {\rm RA} < 304.81$ and $-46.33 < {\rm DEC} < -42.33$. Finally, we mask additional areas with low completeness in 2MPZ, determined by comparing the number counts of 2MPZ sources and 2MASS XSC sources (with $K_s < 13.9$) in NSIDE=64 HEALPixels. We remove pixels with $< 85\%$ completeness, mostly corresponding to areas of lower depth around bright stars. We test variations in the masking procedure (i.e.\ additionally multiplying by the WSC mask, following \cite{Alonso15}, or changing the completeness threshold to 80\% or 90\%) and find minimal changes in results.

\subsection{WSC}

WSC is constructed similarly, but cross-matching AllWISE and SuperCOSMOS only. For WSC, we further apply a color cut of $\rm{W}1-\rm{W}2 >0.2$
to the publicly available sample
to reduce stellar contamination and increase
uniformity \citep{Xavier19}.

For WSC, we follow the masking procedure of \cite{Xavier19}. We start with the mask distributed
with the WSC data release \citep{Bilicki:2016irk}\footnote{\url{http://ssa.roe.ac.uk/WISExSCOSmask.fits.gz}}. We additionally
mask regions with high extinction ($E(B-V) > 0.10$) and high stellar density (density of stars from GAIA greater than 7 times the mean). We additionally test several variations in the masking procedure, adding 
mask at low Galactic latitudes following \cite{RafieiRavandi}; 
adding a WISE bright stars mask \citep{Krolewski20}; and adding a mask of regions in WISE with high moon contamination,
as determined by HEALPix pixels in which GLADE+ \citep{Dalya:2021ewn} is incomplete compared to WSC.  We also test variations
in the sample-selection procedure, i.e.\ additionally using the SVM catalog of \cite{Krakowski16} to restrict
to likely galaxies \citep{RafieiRavandi, Novaes}.  We find that these variations generally lead to a scale-independent shift in the amplitude of $C_{\ell}^{gg}$, either corresponding to a change in galaxy bias
due to differing populations or a change in the stellar contamination fraction, which is entirely degenerate with bias at $\ell > 10$ where the stellar power spectrum is small. In this regime, the effect of changing
stellar contamination is degenerate with bias in both the galaxy auto spectrum and the galaxy cross-spectrum
with GW sources, so it will not cause systematic errors in our modeling.

\section{Validation of the cross-correlation pipeline}\label{validation}
\begin{figure}
    \centering
    \includegraphics[width=0.8\linewidth]{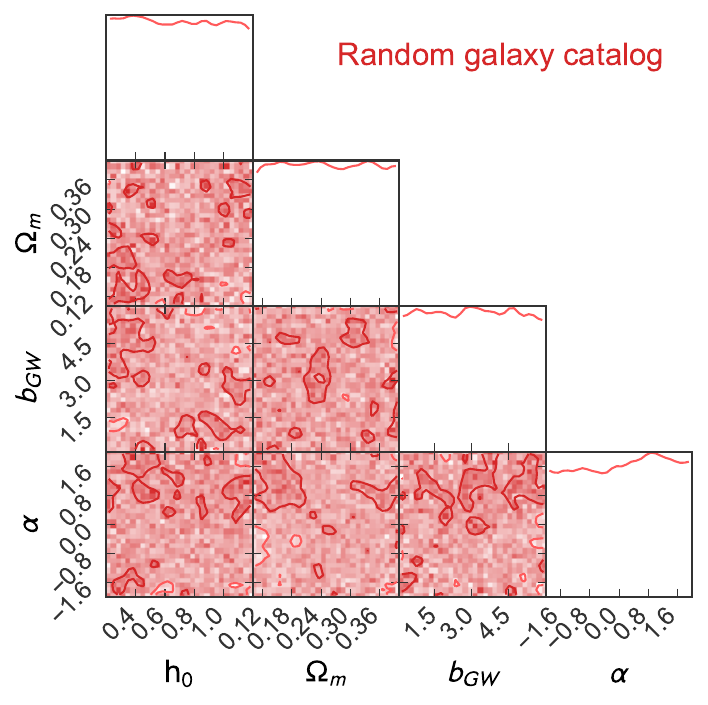}
    \caption{{The constraints on  the value of $H_0$ obtained by cross-correlating with a random galaxy catalog having no spatial correlation and with a galaxy distribution $n(z)$  matching 2MPZ and WSC.}}
    \label{fig:validation}
\end{figure}

To validate the cross-correlation pipeline, we apply our method to a randomly distributed galaxy catalog having no spatial correlations. As a result, the GW sources in the GWTC-3 will also not exhibit any spatial correlation with these galaxies. The randomly generated galaxy catalog has a comoving number distribution $n(z)$ matching the galaxy distribution in 2MPZ and WSC. We show the plot for the cross-correlation on $H_0, \Omega_m$, and the bias parameter in Fig. \ref{fig:validation}. The plot indicates no constraints on the value of the Hubble constant with a random catalog. We have also tested our pipeline with a constant $n(z)$ random catalog and it shows no constraints on the value of the Hubble constant as well.

\bibliography{APJ-H0}{}

\begin{thebibliography}{}
\expandafter\ifx\csname natexlab\endcsname\relax\def\natexlab#1{#1}\fi
\providecommand{\url}[1]{\href{#1}{#1}}
\providecommand{\dodoi}[1]{doi:~\href{http://doi.org/#1}{\nolinkurl{#1}}}
\providecommand{\doeprint}[1]{\href{http://ascl.net/#1}{\nolinkurl{http://ascl.net/#1}}}
\providecommand{\doarXiv}[1]{\href{https://arxiv.org/abs/#1}{\nolinkurl{https://arxiv.org/abs/#1}}}

\bibitem[{Aasi {et~al.}(2015)}]{LIGOScientific:2014pky}
Aasi, J., {et~al.} 2015, Class. Quant. Grav., 32, 074001,
  \dodoi{10.1088/0264-9381/32/7/074001}

\bibitem[{Abbott {et~al.}(2019)}]{Abbott:2019yzh}
Abbott, B., {et~al.} 2019, arXiv:1908.06060.
\newblock \doarXiv{1908.06060}

\bibitem[{Abbott {et~al.}(2016{\natexlab{a}})Abbott, Abbott, Abbott, Abernathy,
  Acernese, Ackley, Adams, Adams, Addesso, Adhikari, Adya, Affeldt, Agathos,
  Agatsuma, Aggarwal, Aguiar, Aiello, Ain, Ajith, Allen, Allocca, Altin,
  Anderson, Anderson, Arai, Arain, Araya, Arceneaux, Areeda, Arnaud, Arun,
  Ascenzi, Ashton, Ast, Aston, Astone, Aufmuth, Aulbert, Babak, Bacon, Bader,
  Baker, Baldaccini, Ballardin, Ballmer, Barayoga, Barclay, Barish, Barker,
  Barone, Barr, Barsotti, Barsuglia, Barta, Bartlett, Barton, Bartos, Bassiri,
  Basti, Batch, Baune, Bavigadda, Bazzan, Behnke, Bejger, Belczynski, Bell,
  Bell, Berger, Bergman, Bergmann, Berry, Bersanetti, Bertolini, Betzwieser,
  Bhagwat, Bhandare, Bilenko, Billingsley, Birch, Birney, Birnholtz, Biscans,
  Bisht, Bitossi, Biwer, Bizouard, Blackburn, Blair, Blair, Blair, Bloemen,
  Bock, Bodiya, Boer, Bogaert, Bogan, Bohe, Bojtos, Bond, Bondu, Bonnand, Boom,
  Bork, Boschi, Bose, Bouffanais, Bozzi, Bradaschia, Brady, Braginsky,
  Branchesi, Brau, Briant, Brillet, Brinkmann, Brisson, Brockill, Brooks,
  Brown, Brown, Brown, Buchanan, Buikema, Bulik, Bulten, Buonanno, Buskulic,
  Buy, Byer, Cabero, Cadonati, Cagnoli, Cahillane, Bustillo, Callister,
  Calloni, Camp, Cannon, Cao, Capano, Capocasa, Carbognani, Caride, Diaz,
  Casentini, Caudill, Cavagli\`a, Cavalier, Cavalieri, Cella, Cepeda, Baiardi,
  Cerretani, Cesarini, Chakraborty, Chalermsongsak, Chamberlin, Chan, Chao,
  Charlton, Chassande-Mottin, Chen, Chen, Cheng, Chincarini, Chiummo, Cho, Cho,
  Chow, Christensen, Chu, Chua, Chung, Ciani, Clara, Clark, Cleva, Coccia,
  Cohadon, Colla, Collette, Cominsky, Constancio, Conte, Conti, Cook, Corbitt,
  Cornish, Corsi, Cortese, Costa, Coughlin, Coughlin, Coulon, Countryman,
  Couvares, Cowan, Coward, Cowart, Coyne, Coyne, Craig, Creighton, Creighton,
  Cripe, Crowder, Cruise, Cumming, Cunningham, Cuoco, Canton, Danilishin,
  D'Antonio, Danzmann, Darman, Da~Silva~Costa, Dattilo, Dave, Daveloza, Davier,
  Davies, Daw, Day, De, DeBra, Debreczeni, Degallaix, De~Laurentis,
  Del\'eglise, Del~Pozzo, Denker, Dent, Dereli, Dergachev, DeRosa, De~Rosa,
  DeSalvo, Dhurandhar, D\'{\i}az, Di~Fiore, Di~Giovanni, Di~Lieto, Di~Pace,
  Di~Palma, Di~Virgilio, Dojcinoski, Dolique, Donovan, Dooley, Doravari,
  Douglas, Downes, Drago, Drever, Driggers, Du, Ducrot, Dwyer, Edo, Edwards,
  Effler, Eggenstein, Ehrens, Eichholz, Eikenberry, Engels, Essick, Etzel,
  Evans, Evans, Everett, Factourovich, Fafone, Fair, Fairhurst, Fan, Fang,
  Farinon, Farr, Farr, Favata, Fays, Fehrmann, Fejer, Feldbaum, Ferrante,
  Ferreira, Ferrini, Fidecaro, Finn, Fiori, Fiorucci, Fisher, Flaminio,
  Fletcher, Fong, Fournier, Franco, Frasca, Frasconi, Frede, Frei, Freise,
  Frey, Frey, Fricke, Fritschel, Frolov, Fulda, Fyffe, Gabbard, Gair,
  Gammaitoni, Gaonkar, Garufi, Gatto, Gaur, Gehrels, Gemme, Gendre, Genin,
  Gennai, George, Gergely, Germain, Ghosh, Ghosh, Ghosh, Giaime, Giardina,
  Giazotto, Gill, Glaefke, Gleason, Goetz, Goetz, Gondan, Gonz\'alez, Castro,
  Gopakumar, Gordon, Gorodetsky, Gossan, Gosselin, Gouaty, Graef, Graff,
  Granata, Grant, Gras, Gray, Greco, Green, Greenhalgh, Groot, Grote,
  Grunewald, Guidi, Guo, Gupta, Gupta, Gushwa, Gustafson, Gustafson, Hacker,
  Hall, Hall, Hammond, Haney, Hanke, Hanks, Hanna, Hannam, Hanson, Hardwick,
  Harms, Harry, Harry, Hart, Hartman, Haster, Haughian, Healy, Heefner,
  Heidmann, Heintze, Heinzel, Heitmann, Hello, Hemming, Hendry, Heng, Hennig,
  Heptonstall, Heurs, Hild, Hoak, Hodge, Hofman, Hollitt, Holt, Holz, Hopkins,
  Hosken, Hough, Houston, Howell, Hu, Huang, Huerta, Huet, Hughey, Husa,
  Huttner, Huynh-Dinh, Idrisy, Indik, Ingram, Inta, Isa, Isac, Isi, Islas,
  Isogai, Iyer, Izumi, Jacobson, Jacqmin, Jang, Jani, Jaranowski, Jawahar,
  Jim\'enez-Forteza, Johnson, Johnson-McDaniel, Jones, Jones, Jonker, Ju,
  Haris, Kalaghatgi, Kalogera, Kandhasamy, Kang, Kanner, Karki, Kasprzack,
  Katsavounidis, Katzman, Kaufer, Kaur, Kawabe, Kawazoe, K\'ef\'elian, Kehl,
  Keitel, Kelley, Kells, Kennedy, Keppel, Key, Khalaidovski, Khalili, Khan,
  Khan, Khan, Khazanov, Kijbunchoo, Kim, Kim, Kim, Kim, Kim, Kim, King, King,
  Kinzel, Kissel, Kleybolte, Klimenko, Koehlenbeck, Kokeyama, Koley,
  Kondrashov, Kontos, Koranda, Korobko, Korth, Kowalska, Kozak, Kringel,
  Krishnan, Kr\'olak, Krueger, Kuehn, Kumar, Kumar, Kuo, Kutynia, Kwee, Lackey,
  Landry, Lange, Lantz, Lasky, Lazzarini, Lazzaro, Leaci, Leavey, Lebigot, Lee,
  Lee, Lee, Lee, Lenon, Leonardi, Leong, Leroy, Letendre, Levin, Levine, Li,
  Libson, Littenberg, Lockerbie, Logue, Lombardi, London, Lord, Lorenzini,
  Loriette, Lormand, Losurdo, Lough, Lousto, Lovelace, L\"uck, Lundgren, Luo,
  Lynch, Ma, MacDonald, Machenschalk, MacInnis, Macleod, Maga\~na Sandoval,
  Magee, Mageswaran, Majorana, Maksimovic, Malvezzi, Man, Mandel, Mandic,
  Mangano, Mansell, Manske, Mantovani, Marchesoni, Marion, M\'arka, M\'arka,
  Markosyan, Maros, Martelli, Martellini, Martin, Martin, Martynov, Marx,
  Mason, Masserot, Massinger, Masso-Reid, Matichard, Matone, Mavalvala,
  Mazumder, Mazzolo, McCarthy, McClelland, McCormick, McGuire, McIntyre,
  McIver, McManus, McWilliams, Meacher, Meadors, Meidam, Melatos, Mendell,
  Mendoza-Gandara, Mercer, Merilh, Merzougui, Meshkov, Messenger, Messick,
  Meyers, Mezzani, Miao, Michel, Middleton, Mikhailov, Milano, Miller,
  Millhouse, Minenkov, Ming, Mirshekari, Mishra, Mitra, Mitrofanov,
  Mitselmakher, Mittleman, Moggi, Mohan, Mohapatra, Montani, Moore, Moore,
  Moraru, Moreno, Morriss, Mossavi, Mours, Mow-Lowry, Mueller, Mueller, Muir,
  Mukherjee, Mukherjee, Mukherjee, Mukund, Mullavey, Munch, Murphy, Murray,
  Mytidis, Nardecchia, Naticchioni, Nayak, Necula, Nedkova, Nelemans, Neri,
  Neunzert, Newton, Nguyen, Nielsen, Nissanke, Nitz, Nocera, Nolting,
  Normandin, Nuttall, Oberling, Ochsner, O'Dell, Oelker, Ogin, Oh, Oh, Ohme,
  Oliver, Oppermann, Oram, O'Reilly, O'Shaughnessy, Ott, Ottaway, Ottens,
  Overmier, Owen, Pai, Pai, Palamos, Palashov, Palomba, Pal-Singh, Pan, Pan,
  Pankow, Pannarale, Pant, Paoletti, Paoli, Papa, Paris, Parker, Pascucci,
  Pasqualetti, Passaquieti, Passuello, Patricelli, Patrick, Pearlstone,
  Pedraza, Pedurand, Pekowsky, Pele, Penn, Perreca, Pfeiffer, Phelps, Piccinni,
  Pichot, Pickenpack, Piergiovanni, Pierro, Pillant, Pinard, Pinto, Pitkin,
  Poeld, Poggiani, Popolizio, Post, Powell, Prasad, Predoi, Premachandra,
  Prestegard, Price, Prijatelj, Principe, Privitera, Prix, Prodi, Prokhorov,
  Puncken, Punturo, Puppo, P\"urrer, Qi, Qin, Quetschke, Quintero,
  Quitzow-James, Raab, Rabeling, Radkins, Raffai, Raja, Rakhmanov, Ramet,
  Rapagnani, Raymond, Razzano, Re, Read, Reed, Regimbau, Rei, Reid, Reitze,
  Rew, Reyes, Ricci, Riles, Robertson, Robie, Robinet, Rocchi, Rolland,
  Rollins, Roma, Romano, Romano, Romanov, Romie, Rosi\ifmmode~\acute{n}\else
  \'{n}\fi{}ska, Rowan, R\"udiger, Ruggi, Ryan, Sachdev, Sadecki, Sadeghian,
  Salconi, Saleem, Salemi, Samajdar, Sammut, Sampson, Sanchez, Sandberg,
  Sandeen, Sanders, Sanders, Sassolas, Sathyaprakash, Saulson, Sauter, Savage,
  Sawadsky, Schale, Schilling, Schmidt, Schmidt, Schnabel, Schofield,
  Sch\"onbeck, Schreiber, Schuette, Schutz, Scott, Scott, Sellers, Sengupta,
  Sentenac, Sequino, Sergeev, Serna, Setyawati, Sevigny, Shaddock, Shaffer,
  Shah, Shahriar, Shaltev, Shao, Shapiro, Shawhan, Sheperd, Shoemaker,
  Shoemaker, Siellez, Siemens, Sigg, Silva, Simakov, Singer, Singer, Singh,
  Singh, Singhal, Sintes, Slagmolen, Smith, Smith, Smith, Smith, Son, Sorazu,
  Sorrentino, Souradeep, Srivastava, Staley, Steinke, Steinlechner,
  Steinlechner, Steinmeyer, Stephens, Stevenson, Stone, Strain, Straniero,
  Stratta, Strauss, Strigin, Sturani, Stuver, Summerscales, Sun, Sutton,
  Swinkels, Szczepa\ifmmode~\acute{n}\else \'{n}\fi{}czyk, Tacca, Talukder,
  Tanner, T\'apai, Tarabrin, Taracchini, Taylor, Theeg, Thirugnanasambandam,
  Thomas, Thomas, Thomas, Thorne, Thorne, Thrane, Tiwari, Tiwari, Tokmakov,
  Tomlinson, Tonelli, Torres, Torrie, T\"oyr\"a, Travasso, Traylor, Trifir\`o,
  Tringali, Trozzo, Tse, Turconi, Tuyenbayev, Ugolini, Unnikrishnan, Urban,
  Usman, Vahlbruch, Vajente, Valdes, Vallisneri, van Bakel, van Beuzekom,
  van~den Brand, Van Den~Broeck, Vander-Hyde, van~der Schaaf, van Heijningen,
  van Veggel, Vardaro, Vass, Vas\'uth, Vaulin, Vecchio, Vedovato, Veitch,
  Veitch, Venkateswara, Verkindt, Vetrano, Vicer\'e, Vinciguerra, Vine, Vinet,
  Vitale, Vo, Vocca, Vorvick, Voss, Vousden, Vyatchanin, Wade, Wade, Wade,
  Waldman, Walker, Wallace, Walsh, Wang, Wang, Wang, Wang, Wang, Ward, Ward,
  Warner, Was, Weaver, Wei, Weinert, Weinstein, Weiss, Welborn, Wen,
  We\ss{}els, Westphal, Wette, Whelan, Whitcomb, White, Whiting, Wiesner,
  Wilkinson, Willems, Williams, Williams, Williamson, Willis, Willke, Wimmer,
  Winkelmann, Winkler, Wipf, Wiseman, Wittel, Woan, Worden, Wright, Wu, Yablon,
  Yakushin, Yam, Yamamoto, Yancey, Yap, Yu, Yvert, Zadro\ifmmode~\dot{z}\else
  \.{z}\fi{}ny, Zangrando, Zanolin, Zendri, Zevin, Zhang, Zhang, Zhang, Zhang,
  Zhao, Zhou, Zhou, Zhu, Zucker, Zuraw, \& Zweizig}]{PhysRevLett.116.061102}
Abbott, B.~P., Abbott, R., Abbott, T.~D., {et~al.} 2016{\natexlab{a}}, Phys.
  Rev. Lett., 116, 061102, \dodoi{10.1103/PhysRevLett.116.061102}

\bibitem[{Abbott {et~al.}(2016{\natexlab{b}})}]{Martynov:2016fzi}
Abbott, B.~P., {et~al.} 2016{\natexlab{b}}, Phys. Rev. D, 93, 112004,
  \dodoi{10.1103/PhysRevD.93.112004}

\bibitem[{Abbott {et~al.}(2020)}]{Abbott:2020khf}
Abbott, R., {et~al.} 2020, Astrophys. J., 896, L44,
  \dodoi{10.3847/2041-8213/ab960f}

\bibitem[{Abbott {et~al.}(2021{\natexlab{a}})}]{LIGOScientific:2021aug}
---. 2021{\natexlab{a}}.
\newblock \doarXiv{2111.03604}

\bibitem[{Abbott {et~al.}(2021{\natexlab{b}})}]{LIGOScientific:2021djp}
---. 2021{\natexlab{b}}.
\newblock \doarXiv{2111.03606}

\bibitem[{Acernese {et~al.}(2014)Acernese, Agathos, Agatsuma, Aisa, Allemandou,
  Allocca, Amarni, Astone, Balestri, Ballardin, \& et~al.}]{Acernese_2014}
Acernese, F., Agathos, M., Agatsuma, K., {et~al.} 2014, Classical and Quantum
  Gravity, 32, 024001, \dodoi{10.1088/0264-9381/32/2/024001}

\bibitem[{Acernese {et~al.}(2019)Acernese, Agathos, Aiello, Allocca, Amato,
  Ansoldi, Antier, Ar\`ene, Arnaud, Ascenzi, Astone, Aubin, Babak, Bacon,
  Badaracco, Bader, Baird, Baldaccini, Ballardin, Baltus, Barbieri, Barneo,
  Barone, Barsuglia, Barta, Basti, Bawaj, Bazzan, Bejger, Belahcene, Bernuzzi,
  Bersanetti, Bertolini, Bischi, Bitossi, Bizouard, Bobba, Boer, Bogaert,
  Bondu, Bonnand, Boom, Boschi, Bouffanais, Bozzi, Bradaschia, Branchesi,
  Breschi, Briant, Brighenti, Brillet, Brooks, Bruno, Bulik, Bulten, Buskulic,
  Cagnoli, Calloni, Canepa, Carapella, Carbognani, Carullo, Casanueva~Diaz,
  Casentini, Casta\~neda, Caudill, Cavalier, Cavalieri, Cella, Cerd\'a-Dur\'an,
  Cesarini, Chaibi, Chassande-Mottin, Chiadini, Chierici, Chincarini, Chiummo,
  Christensen, Chua, Ciani, Ciecielag, Cie\ifmmode~\acute{s}\else
  \'{s}\fi{}lar, Ciolfi, Cipriano, Cirone, Clesse, Cleva, Coccia, Cohadon,
  Cohen, Colpi, Conti, Cordero-Carri\'on, Corezzi, Corre, Cortese, Coulon,
  Croquette, Cudell, Cuoco, Curylo, D'Angelo, D'Antonio, Dattilo, Davier,
  Degallaix, De~Laurentis, Del\'eglise, Del~Pozzo, De~Pietri, De~Rosa,
  De~Rossi, Dietrich, Di~Fiore, Di~Giorgio, Di~Giovanni, Di~Giovanni,
  Di~Girolamo, Di~Lieto, Di~Pace, Di~Palma, Di~Renzo, Drago, Ducoin, Durante,
  D'Urso, Eisenmann, Errico, Estevez, Fafone, Farinon, Feng, Fenyvesi,
  Ferrante, Fidecaro, Fiori, Fiorucci, Fittipaldi, Fiumara, Flaminio, Font,
  Fournier, Frasca, Frasconi, Frey, Fronz\`e, Garufi, Gemme, Genin, Gennai,
  Ghosh, Giacomazzo, Gosselin, Gouaty, Grado, Granata, Greco, Grignani,
  Grimaldi, Grimm, Gruning, Guidi, Guix\'e, Guo, Gupta, Halim, Harder, Harms,
  Heidmann, Heitmann, Hello, Hemming, Hennes, Hinderer, Hofman, Huet, Hui,
  Idzkowski, Iess, Intini, Isac, Jacqmin, Jaranowski, Jonker, Katsanevas,
  K\'ef\'elian, Khan, Khetan, Koekoek, Koley, Kr\'olak, Kutynia, Laghi,
  Lamberts, La~Rosa, Lartaux-Vollard, Lazzaro, Leaci, Leroy, Letendre, Linde,
  Llorens-Monteagudo, Longo, Lorenzini, Loriette, Losurdo, Lumaca, Macquet,
  Majorana, Maksimovic, Man, Mangano, Mantovani, Mapelli, Marchesoni, Marion,
  Marquina, Marsat, Martelli, Martinez, Masserot, Mastrogiovanni, Mejuto~Villa,
  Mereni, Merzougui, Metzdorff, Miani, Michel, Milano, Miller, Milotti,
  Minazzoli, Minenkov, Montani, Morawski, Mours, Muciaccia, Nagar, Nardecchia,
  Naticchioni, Neilson, Nelemans, Nguyen, Nichols, Nissanke, Nocera, Oganesyan,
  Olivetto, Pagano, Pagliaroli, Palomba, Pang, Pannarale, Paoletti, Paoli,
  Pascucci, Pasqualetti, Passaquieti, Passuello, Patricelli, Perego, Pegoraro,
  P\'erigois, Perreca, Perri\`es, Phukon, Piccinni, Pichot, Piendibene,
  Piergiovanni, Pierro, Pillant, Pinard, Pinto, Piotrzkowski, Plastino,
  Poggiani, Popolizio, Porter, Prevedelli, Principe, Prodi, Punturo, Puppo,
  Raaijmakers, Radulesco, Rapagnani, Razzano, Regimbau, Rei, Rettegno, Ricci,
  Riemenschneider, Robinet, Rocchi, Rolland, Romanelli, Romano,
  Rosi\ifmmode~\acute{n}\else \'{n}\fi{}ska, Ruggi, Salafia, Salconi, Samajdar,
  Sanchis-Gual, Santos, Sassolas, Sauter, Sayah, Sentenac, Sequino, Sharma,
  Sieniawska, Singh, Singhal, Sipala, Sordini, Sorrentino, Spera, Stachie,
  Steer, Stratta, Sur, Swinkels, Tacca, Tanasijczuk, Tapia San~Martin, Tiwari,
  Tonelli, Torres-Forn\'e, Tosta~e Melo, Travasso, Tringali, Trovato, Tsang,
  Turconi, Valentini, van Bakel, van Beuzekom, van~den Brand, Van Den~Broeck,
  van~der Schaaf, Vardaro, Vas\'uth, Vedovato, Verkindt, Vetrano, Vicer\'e,
  Vinet, Vocca, Walet, Was, Zadro\ifmmode~\dot{z}\else \.{z}\fi{}ny, Zelenova,
  Zendri, Vahlbruch, Mehmet, L\"uck, \& Danzmann}]{PhysRevLett.123.231108}
Acernese, F., Agathos, M., Aiello, L., {et~al.} 2019, Phys. Rev. Lett., 123,
  231108, \dodoi{10.1103/PhysRevLett.123.231108}

\bibitem[{Aghamousa {et~al.}(2016)}]{Aghamousa:2016zmz}
Aghamousa, A., {et~al.} 2016.
\newblock \doarXiv{1611.00036}

\bibitem[{Aghanim {et~al.}(2018)}]{Aghanim:2018eyx}
Aghanim, N., {et~al.} 2018, arXiv: 1807.06209.
\newblock \doarXiv{1807.06209}

\bibitem[{Aghanim {et~al.}(2020)}]{Planck:2018vyg}
---. 2020, Astron. Astrophys., 641, A6, \dodoi{10.1051/0004-6361/201833910}

\bibitem[{Akutsu {et~al.}(2019)}]{Akutsu:2018axf}
Akutsu, T., {et~al.} 2019, Nat. Astron., 3, 35,
  \dodoi{10.1038/s41550-018-0658-y}

\bibitem[{Akutsu {et~al.}(2020)}]{KAGRA:2020tym}
---. 2020, arXiv:2005.05574.
\newblock \doarXiv{2005.05574}

\bibitem[{{Alonso} {et~al.}(2015){Alonso}, {Salvador}, {S{\'a}nchez},
  {Bilicki}, {Garc{\'\i}a-Bellido}, \& {S{\'a}nchez}}]{Alonso15}
{Alonso}, D., {Salvador}, A.~I., {S{\'a}nchez}, F.~J., {et~al.} 2015, \mnras,
  449, 670, \dodoi{10.1093/mnras/stv309}

\bibitem[{Alonso {et~al.}(2019)Alonso, Sanchez, \& Slosar}]{Alonso:2018jzx}
Alonso, D., Sanchez, J., \& Slosar, A. 2019, Mon. Not. Roy. Astron. Soc., 484,
  4127, \dodoi{10.1093/mnras/stz093}

\bibitem[{{Astropy Collaboration} {et~al.}(2013){Astropy Collaboration},
  {Robitaille}, {Tollerud}, {Greenfield}, {Droettboom}, {Bray}, {Aldcroft},
  {Davis}, {Ginsburg}, {Price-Whelan}, {Kerzendorf}, {Conley}, {Crighton},
  {Barbary}, {Muna}, {Ferguson}, {Grollier}, {Parikh}, {Nair}, {Unther},
  {Deil}, {Woillez}, {Conseil}, {Kramer}, {Turner}, {Singer}, {Fox}, {Weaver},
  {Zabalza}, {Edwards}, {Azalee Bostroem}, {Burke}, {Casey}, {Crawford},
  {Dencheva}, {Ely}, {Jenness}, {Labrie}, {Lim}, {Pierfederici}, {Pontzen},
  {Ptak}, {Refsdal}, {Servillat}, \& {Streicher}}]{2013A&A...558A..33A}
{Astropy Collaboration}, {Robitaille}, T.~P., {Tollerud}, E.~J., {et~al.} 2013,
  \aap, 558, A33, \dodoi{10.1051/0004-6361/201322068}

\bibitem[{{Astropy Collaboration} {et~al.}(2018){Astropy Collaboration},
  {Price-Whelan}, {Sip{\H{o}}cz}, {G{\"u}nther}, {Lim}, {Crawford}, {Conseil},
  {Shupe}, {Craig}, {Dencheva}, {Ginsburg}, {VanderPlas}, {Bradley},
  {P{\'e}rez-Su{\'a}rez}, {de Val-Borro}, {Aldcroft}, {Cruz}, {Robitaille},
  {Tollerud}, {Ardelean}, {Babej}, {Bach}, {Bachetti}, {Bakanov}, {Bamford},
  {Barentsen}, {Barmby}, {Baumbach}, {Berry}, {Biscani}, {Boquien}, {Bostroem},
  {Bouma}, {Brammer}, {Bray}, {Breytenbach}, {Buddelmeijer}, {Burke},
  {Calderone}, {Cano Rodr{\'\i}guez}, {Cara}, {Cardoso}, {Cheedella}, {Copin},
  {Corrales}, {Crichton}, {D'Avella}, {Deil}, {Depagne}, {Dietrich}, {Donath},
  {Droettboom}, {Earl}, {Erben}, {Fabbro}, {Ferreira}, {Finethy}, {Fox},
  {Garrison}, {Gibbons}, {Goldstein}, {Gommers}, {Greco}, {Greenfield},
  {Groener}, {Grollier}, {Hagen}, {Hirst}, {Homeier}, {Horton}, {Hosseinzadeh},
  {Hu}, {Hunkeler}, {Ivezi{\'c}}, {Jain}, {Jenness}, {Kanarek}, {Kendrew},
  {Kern}, {Kerzendorf}, {Khvalko}, {King}, {Kirkby}, {Kulkarni}, {Kumar},
  {Lee}, {Lenz}, {Littlefair}, {Ma}, {Macleod}, {Mastropietro}, {McCully},
  {Montagnac}, {Morris}, {Mueller}, {Mumford}, {Muna}, {Murphy}, {Nelson},
  {Nguyen}, {Ninan}, {N{\"o}the}, {Ogaz}, {Oh}, {Parejko}, {Parley}, {Pascual},
  {Patil}, {Patil}, {Plunkett}, {Prochaska}, {Rastogi}, {Reddy Janga},
  {Sabater}, {Sakurikar}, {Seifert}, {Sherbert}, {Sherwood-Taylor}, {Shih},
  {Sick}, {Silbiger}, {Singanamalla}, {Singer}, {Sladen}, {Sooley},
  {Sornarajah}, {Streicher}, {Teuben}, {Thomas}, {Tremblay}, {Turner},
  {Terr{\'o}n}, {van Kerkwijk}, {de la Vega}, {Watkins}, {Weaver}, {Whitmore},
  {Woillez}, {Zabalza}, \& {Astropy Contributors}}]{2018AJ....156..123A}
{Astropy Collaboration}, {Price-Whelan}, A.~M., {Sip{\H{o}}cz}, B.~M., {et~al.}
  2018, \aj, 156, 123, \dodoi{10.3847/1538-3881/aabc4f}

\bibitem[{{Balaguera-Antol{\'\i}nez} {et~al.}(2018){Balaguera-Antol{\'\i}nez},
  {Bilicki}, {Branchini}, \& {Postiglione}}]{Balaguera-Antolinez}
{Balaguera-Antol{\'\i}nez}, A., {Bilicki}, M., {Branchini}, E., \&
  {Postiglione}, A. 2018, \mnras, 476, 1050, \dodoi{10.1093/mnras/sty262}

\bibitem[{Bera {et~al.}(2020)Bera, Rana, More, \& Bose}]{Bera:2020jhx}
Bera, S., Rana, D., More, S., \& Bose, S. 2020, Astrophys. J., 902, 79,
  \dodoi{10.3847/1538-4357/abb4e0}

\bibitem[{Bilicki {et~al.}(2014)Bilicki, Jarrett, Peacock, Cluver, \&
  Steward}]{Bilicki:2013sza}
Bilicki, M., Jarrett, T.~H., Peacock, J.~A., Cluver, M.~E., \& Steward, L.
  2014, Astrophys. J. Suppl., 210, 9, \dodoi{10.1088/0067-0049/210/1/9}

\bibitem[{Bilicki {et~al.}(2016)}]{Bilicki:2016irk}
Bilicki, M., {et~al.} 2016, Astrophys. J. Suppl., 225, 5,
  \dodoi{10.3847/0067-0049/225/1/5}

\bibitem[{Blanchard {et~al.}(2020)}]{Euclid:2019clj}
Blanchard, A., {et~al.} 2020, Astron. Astrophys., 642, A191,
  \dodoi{10.1051/0004-6361/202038071}

\bibitem[{Bocquet \& Carter(2016)}]{Bocquet2016}
Bocquet, S., \& Carter, F.~W. 2016, The Journal of Open Source Software, 1,
  \dodoi{10.21105/joss.00046}

\bibitem[{Brown {et~al.}(2016)Brown, Vallenari, Prusti, de~Bruijne, Mignard,
  Drimmel, Babusiaux, Bailer-Jones, Bastian, Biermann, Evans, Eyer, Jansen,
  Jordi, Katz, Klioner, Lammers, Lindegren, Luri, O’Mullane, Panem, Pourbaix,
  Randich, Sartoretti, Siddiqui, Soubiran, Valette, van Leeuwen, Walton, Aerts,
  Arenou, Cropper, Høg, Lattanzi, Grebel, Holland, Huc, Passot, Perryman,
  Bramante, Cacciari, Castañeda, Chaoul, Cheek, De~Angeli, Fabricius, Guerra,
  Hernández, Jean-Antoine-Piccolo, Masana, Messineo, Mowlavi, Nienartowicz,
  Ordóñez-Blanco, Panuzzo, Portell, Richards, Riello, Seabroke, Tanga,
  Thévenin, Torra, Els, Gracia-Abril, Comoretto, Garcia-Reinaldos, Lock,
  Mercier, Altmann, Andrae, Astraatmadja, Bellas-Velidis, Benson, Berthier,
  Blomme, Busso, Carry, Cellino, Clementini, Cowell, Creevey, Cuypers,
  Davidson, De~Ridder, de~Torres, Delchambre, Dell’Oro, Ducourant, Frémat,
  García-Torres, Gosset, Halbwachs, Hambly, Harrison, Hauser, Hestroffer,
  Hodgkin, Huckle, Hutton, Jasniewicz, Jordan, Kontizas, Korn, Lanzafame,
  Manteiga, Moitinho, Muinonen, Osinde, Pancino, Pauwels, Petit, Recio-Blanco,
  Robin, Sarro, Siopis, Smith, Smith, Sozzetti, Thuillot, van Reeven, Viala,
  Abbas, Abreu~Aramburu, Accart, Aguado, Allan, Allasia, Altavilla, Álvarez,
  Alves, Anderson, Andrei, Anglada~Varela, Antiche, Antoja, Antón, Arcay,
  Bach, Baker, Balaguer-Núñez, Barache, Barata, Barbier, Barblan, Barrado~y
  Navascués, Barros, Barstow, Becciani, Bellazzini, Bello~García, Belokurov,
  Bendjoya, Berihuete, Bianchi, Bienaymé, Billebaud, Blagorodnova,
  Blanco-Cuaresma, Boch, Bombrun, Borrachero, Bouquillon, Bourda, Bouy,
  Bragaglia, Breddels, Brouillet, Brüsemeister, Bucciarelli, Burgess, Burgon,
  Burlacu, Busonero, Buzzi, Caffau, Cambras, Campbell, Cancelliere,
  Cantat-Gaudin, Carlucci, Carrasco, Castellani, Charlot, Charnas, Chiavassa,
  Clotet, Cocozza, Collins, Costigan, Crifo, Cross, Crosta, Crowley, Dafonte,
  Damerdji, Dapergolas, David, David, De~Cat, de~Felice, de~Laverny, De~Luise,
  De~March, de~Martino, de~Souza, Debosscher, del Pozo, Delbo, Delgado,
  Delgado, Di~Matteo, Diakite, Distefano, Dolding, Dos~Anjos, Drazinos, Duran,
  Dzigan, Edvardsson, Enke, Evans, Eynard~Bontemps, Fabre, Fabrizio, Faigler,
  Falcão, Farràs~Casas, Federici, Fedorets, Fernández-Hernández, Fernique,
  Fienga, Figueras, Filippi, Findeisen, Fonti, Fouesneau, Fraile, Fraser,
  Fuchs, Gai, Galleti, Galluccio, Garabato, García-Sedano, Garofalo, Garralda,
  Gavras, Gerssen, Geyer, Gilmore, Girona, Giuffrida, Gomes, González-Marcos,
  González-Núñez, González-Vidal, Granvik, Guerrier, Guillout, Guiraud,
  Gúrpide, Gutiérrez-Sánchez, Guy, Haigron, Hatzidimitriou, Haywood, Heiter,
  Helmi, Hobbs, Hofmann, Holl, Holland, Hunt, Hypki, Icardi, Irwin, Jevardat~de
  Fombelle, Jofré, Jonker, Jorissen, Julbe, Karampelas, Kochoska, Kohley,
  Kolenberg, Kontizas, Koposov, Kordopatis, Koubsky, Krone-Martins,
  Kudryashova, Kull, Bachchan, Lacoste-Seris, Lanza, Lavigne,
  Le~Poncin-Lafitte, Lebreton, Lebzelter, Leccia, Leclerc, Lecoeur-Taibi,
  Lemaitre, Lenhardt, Leroux, Liao, Licata, Lindstrøm, Lister, Livanou, Lobel,
  Löffler, López, Lorenz, MacDonald, Magalhães~Fernandes, Managau, Mann,
  Mantelet, Marchal, Marchant, Marconi, Marinoni, Marrese, Marschalkó,
  Marshall, Martín-Fleitas, Martino, Mary, Matijevič, Mazeh, McMillan,
  Messina, Michalik, Millar, Miranda, Molina, Molinaro, Molinaro, Molnár,
  Moniez, Montegriffo, Mor, Mora, Morbidelli, Morel, Morgenthaler, Morris,
  Mulone, Muraveva, Musella, Narbonne, Nelemans, Nicastro, Noval, Ordénovic,
  Ordieres-Meré, Osborne, Pagani, Pagano, Pailler, Palacin, Palaversa,
  Parsons, Pecoraro, Pedrosa, Pentikäinen, Pichon, Piersimoni, Pineau, Plachy,
  Plum, Poujoulet, Prša, Pulone, Ragaini, Rago, Rambaux, Ramos-Lerate,
  Ranalli, Rauw, Read, Regibo, Reylé, Ribeiro, Rimoldini, Ripepi, Riva, Rixon,
  Roelens, Romero-Gómez, Rowell, Royer, Ruiz-Dern, Sadowski,
  Sagristà~Sellés, Sahlmann, Salgado, Salguero, Sarasso, Savietto,
  Schultheis, Sciacca, Segol, Segovia, Segransan, Shih, Smareglia, Smart,
  Solano, Solitro, Sordo, Soria~Nieto, Souchay, Spagna, Spoto, Stampa, Steele,
  Steidelmüller, Stephenson, Stoev, Suess, Süveges, Surdej, Szabados,
  Szegedi-Elek, Tapiador, Taris, Tauran, Taylor, Teixeira, Terrett, Tingley,
  Trager, Turon, Ulla, Utrilla, Valentini, van Elteren, Van~Hemelryck, van
  Leeuwen, Varadi, Vecchiato, Veljanoski, Via, Vicente, Vogt, Voss, Votruba,
  Voutsinas, Walmsley, Weiler, Weingrill, Wevers, Wyrzykowski, Yoldas, Žerjal,
  Zucker, Zurbach, Zwitter, Alecu, Allen, Allende~Prieto, Amorim,
  Anglada-Escudé, Arsenijevic, Azaz, Balm, Beck, Bernstein, Bigot, Bijaoui,
  Blasco, Bonfigli, Bono, Boudreault, Bressan, Brown, Brunet, Bunclark,
  Buonanno, Butkevich, Carret, Carrion, Chemin, Chéreau, Corcione, Darmigny,
  de~Boer, de~Teodoro, de~Zeeuw, Delle~Luche, Domingues, Dubath, Fodor,
  Frézouls, Fries, Fustes, Fyfe, Gallardo, Gallegos, Gardiol, Gebran, Gomboc,
  Gómez, Grux, Gueguen, Heyrovsky, Hoar, Iannicola, Isasi~Parache, Janotto,
  Joliet, Jonckheere, Keil, Kim, Klagyivik, Klar, Knude, Kochukhov, Kolka, Kos,
  Kutka, Lainey, LeBouquin, Liu, Loreggia, Makarov, Marseille, Martayan,
  Martinez-Rubi, Massart, Meynadier, Mignot, Munari, Nguyen, Nordlander,
  Ocvirk, O’Flaherty, Olias~Sanz, Ortiz, Osorio, Oszkiewicz, Ouzounis,
  Palmer, Park, Pasquato, Peltzer, Peralta, Péturaud, Pieniluoma, Pigozzi,
  Poels, Prat, Prod’homme, Raison, Rebordao, Risquez, Rocca-Volmerange,
  Rosen, Ruiz-Fuertes, Russo, Sembay, Serraller~Vizcaino, Short, Siebert,
  Silva, Sinachopoulos, Slezak, Soffel, Sosnowska, Straižys, ter Linden,
  Terrell, Theil, Tiede, Troisi, Tsalmantza, Tur, Vaccari, Vachier, Valles,
  Van~Hamme, Veltz, Virtanen, Wallut, Wichmann, Wilkinson, Ziaeepour, \&
  Zschocke}]{GaiaDR1}
Brown, A. G.~A., Vallenari, A., Prusti, T., {et~al.} 2016, Astronomy \&
  Astrophysics, 595, A2, \dodoi{10.1051/0004-6361/201629512}

\bibitem[{{Cawthon} {et~al.}(2022){Cawthon}, {Elvin-Poole}, {Porredon},
  {Crocce}, {Giannini}, {Gatti}, {Ross}, {Rykoff}, {Carnero Rosell}, {DeRose},
  {Lee}, {Rodriguez-Monroy}, {Amon}, {Bechtol}, {De Vicente}, {Gruen},
  {Morgan}, {Sanchez}, {Sanchez}, {Sevilla-Noarbe}, {Abbott}, {Aguena},
  {Allam}, {Annis}, {Avila}, {Bacon}, {Bertin}, {Brooks}, {Burke}, {Carrasco
  Kind}, {Carretero}, {Castander}, {Choi}, {Costanzi}, {da Costa}, {Pereira},
  {Dawson}, {Desai}, {Diehl}, {Eckert}, {Everett}, {Ferrero}, {Fosalba},
  {Frieman}, {Garc{\'\i}a-Bellido}, {Gaztanaga}, {Gruendl}, {Gschwend},
  {Gutierrez}, {Hinton}, {Hollowood}, {Honscheid}, {Huterer}, {James}, {Kim},
  {Kneib}, {Kuehn}, {Kuropatkin}, {Lahav}, {Lima}, {Lin}, {Maia}, {Melchior},
  {Menanteau}, {Miquel}, {Mohr}, {Muir}, {Myles}, {Palmese}, {Pandey},
  {Paz-Chinch{\'o}n}, {Percival}, {Plazas}, {Roodman}, {Rossi}, {Scarpine},
  {Serrano}, {Smith}, {Soares-Santos}, {Suchyta}, {Swanson}, {Tarle}, {To},
  {Troxel}, {Wilkinson}, \& {DES Collaboration}}]{Cawthon22}
{Cawthon}, R., {Elvin-Poole}, J., {Porredon}, A., {et~al.} 2022, \mnras, 513,
  5517, \dodoi{10.1093/mnras/stac1160}

\bibitem[{{Cigarr{\'a}n D{\'\i}az} \& {Mukherjee}(2022)}]{2022MNRAS.511.2782C}
{Cigarr{\'a}n D{\'\i}az}, C., \& {Mukherjee}, S. 2022, \mnras, 511, 2782,
  \dodoi{10.1093/mnras/stac208}

\bibitem[{{Collister} \& {Lahav}(2004)}]{Collister04}
{Collister}, A.~A., \& {Lahav}, O. 2004, \pasp, 116, 345,
  \dodoi{10.1086/383254}

\bibitem[{Dainotti {et~al.}(2021)Dainotti, De~Simone, Schiavone, Montani,
  Rinaldi, \& Lambiase}]{Dainotti:2021pqg}
Dainotti, M.~G., De~Simone, B., Schiavone, T., {et~al.} 2021, Astrophys. J.,
  912, 150, \dodoi{10.3847/1538-4357/abeb73}

\bibitem[{D\'alya {et~al.}(2018)D\'alya, Galg\'oczi, Dobos, Frei, Heng, Macas,
  Messenger, Raffai, \& de~Souza}]{Dalya:2018cnd}
D\'alya, G., Galg\'oczi, G., Dobos, L., {et~al.} 2018, Mon. Not. Roy. Astron.
  Soc., 479, 2374, \dodoi{10.1093/mnras/sty1703}

\bibitem[{D\'alya {et~al.}(2021)}]{Dalya:2021ewn}
D\'alya, G., {et~al.} 2021.
\newblock \doarXiv{2110.06184}

\bibitem[{Di~Valentino {et~al.}(2021)Di~Valentino, Mena, Pan, Visinelli, Yang,
  Melchiorri, Mota, Riess, \& Silk}]{DiValentino:2021izs}
Di~Valentino, E., Mena, O., Pan, S., {et~al.} 2021, Class. Quant. Grav., 38,
  153001, \dodoi{10.1088/1361-6382/ac086d}

\bibitem[{Dore {et~al.}(2018)}]{Dore:2018kgp}
Dore, O., {et~al.} 2018, arXiv.
\newblock \doarXiv{1805.05489}

\bibitem[{Ezquiaga \& Holz(2022)}]{Ezquiaga:2022zkx}
Ezquiaga, J.~M., \& Holz, D.~E. 2022.
\newblock \doarXiv{2202.08240}

\bibitem[{Farmer {et~al.}(2019)Farmer, Renzo, de~Mink, Marchant, \&
  Justham}]{Farmer}
Farmer, R., Renzo, M., de~Mink, S.~E., Marchant, P., \& Justham, S. 2019, The
  Astrophysical Journal, 887, 53, \dodoi{10.3847/1538-4357/ab518b}

\bibitem[{Farr {et~al.}(2019)Farr, Fishbach, Ye, \& Holz}]{Farr:2019twy}
Farr, W.~M., Fishbach, M., Ye, J., \& Holz, D. 2019, Astrophys. J. Lett., 883,
  L42, \dodoi{10.3847/2041-8213/ab4284}

\bibitem[{Finke {et~al.}(2021)Finke, Foffa, Iacovelli, Maggiore, \&
  Mancarella}]{Finke:2021aom}
Finke, A., Foffa, S., Iacovelli, F., Maggiore, M., \& Mancarella, M. 2021,
  arXiv:2101.12660.
\newblock \doarXiv{2101.12660}

\bibitem[{{Foreman-Mackey} {et~al.}(2013){Foreman-Mackey}, {Hogg}, {Lang}, \&
  {Goodman}}]{2013PASP..125..306F}
{Foreman-Mackey}, D., {Hogg}, D.~W., {Lang}, D., \& {Goodman}, J. 2013, \pasp,
  125, 306, \dodoi{10.1086/670067}

\bibitem[{{Gatti} {et~al.}(2022){Gatti}, {Giannini}, {Bernstein}, {Alarcon},
  {Myles}, {Amon}, {Cawthon}, {Troxel}, {DeRose}, {Everett}, {Ross}, {Rykoff},
  {Elvin-Poole}, {Cordero}, {Harrison}, {Sanchez}, {Prat}, {Gruen}, {Lin},
  {Crocce}, {Rozo}, {Abbott}, {Aguena}, {Allam}, {Annis}, {Avila}, {Bacon},
  {Bertin}, {Brooks}, {Burke}, {Rosell}, {Kind}, {Carretero}, {Castander},
  {Choi}, {Conselice}, {Costanzi}, {Crocce}, {da Costa}, {Pereira}, {Dawson},
  {Desai}, {Diehl}, {Eckert}, {Eifler}, {Evrard}, {Ferrero}, {Flaugher},
  {Fosalba}, {Frieman}, {Garc{\'\i}a-Bellido}, {Gaztanaga}, {Giannantonio},
  {Gruendl}, {Gschwend}, {Hinton}, {Hollowood}, {Honscheid}, {Hoyle},
  {Huterer}, {James}, {Kuehn}, {Kuropatkin}, {Lahav}, {Lima}, {MacCrann},
  {Maia}, {March}, {Marshall}, {Melchior}, {Menanteau}, {Miquel}, {Mohr},
  {Morgan}, {Ogando}, {Palmese}, {Paz-Chinch{\'o}n}, {Percival}, {Plazas},
  {Rodriguez-Monroy}, {Roodman}, {Rossi}, {Samuroff}, {Sanchez}, {Scarpine},
  {Secco}, {Serrano}, {Sevilla-Noarbe}, {Smith}, {Soares-Santos}, {Suchyta},
  {Swanson}, {Tarle}, {Thomas}, {To}, {Varga}, {Weller}, {Wilkinson},
  {Wilkinson}, \& {DES Collaboration}}]{Gatti22}
{Gatti}, M., {Giannini}, G., {Bernstein}, G.~M., {et~al.} 2022, \mnras, 510,
  1223, \dodoi{10.1093/mnras/stab3311}

\bibitem[{G\'orski {et~al.}(2005)G\'orski, Hivon, Banday, Wandelt, Hansen,
  Reinecke, \& Bartelman}]{Gorski:2004by}
G\'orski, K.~M., Hivon, E., Banday, A.~J., {et~al.} 2005, Astrophys. J., 622,
  759, \dodoi{10.1086/427976}

\bibitem[{{Grain} {et~al.}(2009){Grain}, {Tristram}, \& {Stompor}}]{Grain09}
{Grain}, J., {Tristram}, M., \& {Stompor}, R. 2009, \prd, 79, 123515,
  \dodoi{10.1103/PhysRevD.79.123515}

\bibitem[{{Hambly} {et~al.}(2001{\natexlab{a}}){Hambly}, {Davenhall}, {Irwin},
  \& {MacGillivray}}]{Hambly01c}
{Hambly}, N.~C., {Davenhall}, A.~C., {Irwin}, M.~J., \& {MacGillivray}, H.~T.
  2001{\natexlab{a}}, \mnras, 326, 1315,
  \dodoi{10.1111/j.1365-2966.2001.04662.x}

\bibitem[{{Hambly} {et~al.}(2001{\natexlab{b}}){Hambly}, {Irwin}, \&
  {MacGillivray}}]{Hambly01b}
{Hambly}, N.~C., {Irwin}, M.~J., \& {MacGillivray}, H.~T. 2001{\natexlab{b}},
  \mnras, 326, 1295, \dodoi{10.1111/j.1365-2966.2001.04661.x}

\bibitem[{{Hambly} {et~al.}(2001{\natexlab{c}}){Hambly}, {MacGillivray},
  {Read}, {Tritton}, {Thomson}, {Kelly}, {Morgan}, {Smith}, {Driver},
  {Williamson}, {Parker}, {Hawkins}, {Williams}, \& {Lawrence}}]{Hambly01a}
{Hambly}, N.~C., {MacGillivray}, H.~T., {Read}, M.~A., {et~al.}
  2001{\natexlab{c}}, \mnras, 326, 1279,
  \dodoi{10.1111/j.1365-2966.2001.04660.x}

\bibitem[{{Heymans} {et~al.}(2021){Heymans}, {Tr{\"o}ster}, {Asgari}, {Blake},
  {Hildebrandt}, {Joachimi}, {Kuijken}, {Lin}, {S{\'a}nchez}, {van den Busch},
  {Wright}, {Amon}, {Bilicki}, {de Jong}, {Crocce}, {Dvornik}, {Erben},
  {Fortuna}, {Getman}, {Giblin}, {Glazebrook}, {Hoekstra}, {Joudaki},
  {Kannawadi}, {K{\"o}hlinger}, {Lidman}, {Miller}, {Napolitano}, {Parkinson},
  {Schneider}, {Shan}, {Valentijn}, {Verdoes Kleijn}, \&
  {Wolf}}]{2021A&A...646A.140H}
{Heymans}, C., {Tr{\"o}ster}, T., {Asgari}, M., {et~al.} 2021, \aap, 646, A140,
  \dodoi{10.1051/0004-6361/202039063}

\bibitem[{{Hivon} {et~al.}(2002){Hivon}, {G{\'o}rski}, {Netterfield}, {Crill},
  {Prunet}, \& {Hansen}}]{Hivon02}
{Hivon}, E., {G{\'o}rski}, K.~M., {Netterfield}, C.~B., {et~al.} 2002, \apj,
  567, 2, \dodoi{10.1086/338126}

\bibitem[{Hunter(2007)}]{Hunter:2007}
Hunter, J.~D. 2007, Computing In Science \& Engineering, 9, 90,
  \dodoi{10.1109/MCSE.2007.55}

\bibitem[{{Jarrett} {et~al.}(2000){Jarrett}, {Chester}, {Cutri}, {Schneider},
  {Skrutskie}, \& {Huchra}}]{Jarrett00}
{Jarrett}, T.~H., {Chester}, T., {Cutri}, R., {et~al.} 2000, \aj, 119, 2498,
  \dodoi{10.1086/301330}

\bibitem[{Jones {et~al.}(2001--)Jones, Oliphant, Peterson, {et~al.}}]{scipy}
Jones, E., Oliphant, T., Peterson, P., {et~al.} 2001--, {SciPy}: Open source
  scientific tools for {Python}.
\newblock \url{http://www.scipy.org/}

\bibitem[{{Koukoufilippas} {et~al.}(2020){Koukoufilippas}, {Alonso}, {Bilicki},
  \& {Peacock}}]{Koukoufilippas}
{Koukoufilippas}, N., {Alonso}, D., {Bilicki}, M., \& {Peacock}, J.~A. 2020,
  \mnras, 491, 5464, \dodoi{10.1093/mnras/stz3351}

\bibitem[{{Krakowski} {et~al.}(2016){Krakowski}, {Ma{\l}ek}, {Bilicki},
  {Pollo}, {Kurcz}, \& {Krupa}}]{Krakowski16}
{Krakowski}, T., {Ma{\l}ek}, K., {Bilicki}, M., {et~al.} 2016, \aap, 596, A39,
  \dodoi{10.1051/0004-6361/201629165}

\bibitem[{{Krolewski} {et~al.}(2020){Krolewski}, {Ferraro}, {Schlafly}, \&
  {White}}]{Krolewski20}
{Krolewski}, A., {Ferraro}, S., {Schlafly}, E.~F., \& {White}, M. 2020, \jcap,
  2020, 047, \dodoi{10.1088/1475-7516/2020/05/047}

\bibitem[{{LSST Science Collaboration} {et~al.}(2009){LSST Science
  Collaboration}, {Abell}, {Allison}, {Anderson}, {Andrew}, {Angel}, {Armus},
  {Arnett}, {Asztalos}, {Axelrod}, \& et~al.}]{LSSTScienceBook}
{LSST Science Collaboration}, {Abell}, P.~A., {Allison}, J., {et~al.} 2009,
  ArXiv e-prints.
\newblock \doarXiv{0912.0201}

\bibitem[{Mastrogiovanni {et~al.}(2021)Mastrogiovanni, Leyde, Karathanasis,
  Chassande-Mottin, Steer, Gair, Ghosh, Gray, Mukherjee, \&
  Rinaldi}]{Mastrogiovanni:2021wsd}
Mastrogiovanni, S., Leyde, K., Karathanasis, C., {et~al.} 2021,
  arXiv:2103.14663.
\newblock \doarXiv{2103.14663}

\bibitem[{Menard {et~al.}(2013)Menard, Scranton, Schmidt, Morrison, Jeong,
  Budavari, \& Rahman}]{Menard:2013aaa}
Menard, B., Scranton, R., Schmidt, S., {et~al.} 2013, arXiv:1303.4722.
\newblock \doarXiv{1303.4722}

\bibitem[{Mukherjee(2021)}]{Mukherjee:2021rtw}
Mukherjee, S. 2021.
\newblock \doarXiv{2112.10256}

\bibitem[{Mukherjee {et~al.}(2021{\natexlab{a}})Mukherjee, Lavaux, Bouchet,
  Jasche, Wandelt, Nissanke, Leclercq, \& Hotokezaka}]{Mukherjee:2019qmm}
Mukherjee, S., Lavaux, G., Bouchet, F.~R., {et~al.} 2021{\natexlab{a}}, Astron.
  Astrophys., 646, A65, \dodoi{10.1051/0004-6361/201936724}

\bibitem[{Mukherjee \& Wandelt(2018)}]{Mukherjee:2018ebj}
Mukherjee, S., \& Wandelt, B.~D. 2018, arXiv:1808.06615.
\newblock \doarXiv{1808.06615}

\bibitem[{Mukherjee {et~al.}(2021{\natexlab{b}})Mukherjee, Wandelt, Nissanke,
  \& Silvestri}]{Mukherjee:2020hyn}
Mukherjee, S., Wandelt, B.~D., Nissanke, S.~M., \& Silvestri, A.
  2021{\natexlab{b}}, Phys. Rev. D, 103, 043520,
  \dodoi{10.1103/PhysRevD.103.043520}

\bibitem[{Mukherjee {et~al.}(2020)Mukherjee, Wandelt, \&
  Silk}]{Mukherjee:2019wcg}
Mukherjee, S., Wandelt, B.~D., \& Silk, J. 2020, Mon. Not. Roy. Astron. Soc.,
  494, 1956, \dodoi{10.1093/mnras/staa827}

\bibitem[{Mukherjee {et~al.}(2021{\natexlab{c}})Mukherjee, Wandelt, \&
  Silk}]{Mukherjee:2020mha}
---. 2021{\natexlab{c}}, Mon. Not. Roy. Astron. Soc., 502, 1136,
  \dodoi{10.1093/mnras/stab001}

\bibitem[{Newman(2008)}]{Newman:2008mb}
Newman, J.~A. 2008, Astrophys. J., 684, 88, \dodoi{10.1086/589982}

\bibitem[{{Novaes} {et~al.}(2018){Novaes}, {Bernui}, {Xavier}, \&
  {Marques}}]{Novaes}
{Novaes}, C.~P., {Bernui}, A., {Xavier}, H.~S., \& {Marques}, G.~A. 2018,
  \mnras, 478, 3253, \dodoi{10.1093/mnras/sty1265}

\bibitem[{Oguri(2016)}]{PhysRevD.93.083511}
Oguri, M. 2016, Phys. Rev. D, 93, 083511, \dodoi{10.1103/PhysRevD.93.083511}

\bibitem[{Palmese {et~al.}(2021)Palmese, Bom, Mucesh, \&
  Hartley}]{Palmese:2021mjm}
Palmese, A., Bom, C.~R., Mucesh, S., \& Hartley, W.~G. 2021.
\newblock \doarXiv{2111.06445}

\bibitem[{{Peacock} \& {Bilicki}(2018)}]{PeacockBilicki18}
{Peacock}, J.~A., \& {Bilicki}, M. 2018, \mnras, 481, 1133,
  \dodoi{10.1093/mnras/sty2314}

\bibitem[{{Peacock} {et~al.}(2016){Peacock}, {Hambly}, {Bilicki},
  {MacGillivray}, {Miller}, {Read}, \& {Tritton}}]{Peacock16}
{Peacock}, J.~A., {Hambly}, N.~C., {Bilicki}, M., {et~al.} 2016, \mnras, 462,
  2085, \dodoi{10.1093/mnras/stw1818}

\bibitem[{P\'erez \& Granger(2007)}]{PER-GRA:2007}
P\'erez, F., \& Granger, B.~E. 2007, Computing in Science and Engineering, 9,
  21, \dodoi{10.1109/MCSE.2007.53}

\bibitem[{{Rafiei-Ravandi} {et~al.}(2021){Rafiei-Ravandi}, {Smith}, {Li},
  {Masui}, {Josephy}, {Dobbs}, {Lang}, {Bhardwaj}, {Patel}, {Bandura},
  {Berger}, {Boyle}, {Brar}, {Breitman}, {Cassanelli}, {Chawla}, {Adam Dong},
  {Fonseca}, {Gaensler}, {Giri}, {Good}, {Halpern}, {Kaczmarek}, {Kaspi},
  {Leung}, {Lin}, {Mena-Parra}, {Meyers}, {Michilli}, {M{\"u}nchmeyer}, {Ng},
  {Petroff}, {Pleunis}, {Rahman}, {Sanghavi}, {Scholz}, {Shin}, {Stairs},
  {Tendulkar}, {Vanderlinde}, \& {Zwaniga}}]{RafieiRavandi}
{Rafiei-Ravandi}, M., {Smith}, K.~M., {Li}, D., {et~al.} 2021, \apj, 922, 42,
  \dodoi{10.3847/1538-4357/ac1dab}

\bibitem[{{Rau} {et~al.}(2023){Rau}, {Dalal}, {Zhang}, {Li}, {Nishizawa},
  {More}, {Mandelbaum}, {Miyatake}, {Strauss}, \& {Takada}}]{Rau23}
{Rau}, M.~M., {Dalal}, R., {Zhang}, T., {et~al.} 2023, \mnras, 524, 5109,
  \dodoi{10.1093/mnras/stad1962}

\bibitem[{Riess {et~al.}(2019)Riess, Casertano, Yuan, Macri, \&
  Scolnic}]{Riess:2019cxk}
Riess, A.~G., Casertano, S., Yuan, W., Macri, L.~M., \& Scolnic, D. 2019,
  Astrophys. J., 876, 85, \dodoi{10.3847/1538-4357/ab1422}

\bibitem[{Riess {et~al.}(2021)}]{Riess:2021jrx}
Riess, A.~G., {et~al.} 2021.
\newblock \doarXiv{2112.04510}

\bibitem[{{Schlegel} {et~al.}(1998){Schlegel}, {Finkbeiner}, \&
  {Davis}}]{sfd98}
{Schlegel}, D.~J., {Finkbeiner}, D.~P., \& {Davis}, M. 1998, \apj, 500, 525,
  \dodoi{10.1086/305772}

\bibitem[{Schmidt {et~al.}(2013)Schmidt, Menard, Scranton, Morrison, \&
  McBride}]{Schmidt:2013sba}
Schmidt, S., Menard, B., Scranton, R., Morrison, C., \& McBride, C. 2013, Mon.
  Not. Roy. Astron. Soc., 431, 3307, \dodoi{10.1093/mnras/stt410}

\bibitem[{{Schutz}(1986)}]{Schutz}
{Schutz}, B.~F. 1986, \nat, 323, 310, \dodoi{10.1038/323310a0}

\bibitem[{{Skrutskie} {et~al.}(2006){Skrutskie}, {Cutri}, {Stiening},
  {Weinberg}, {Schneider}, {Carpenter}, {Beichman}, {Capps}, {Chester},
  {Elias}, {Huchra}, {Liebert}, {Lonsdale}, {Monet}, {Price}, {Seitzer},
  {Jarrett}, {Kirkpatrick}, {Gizis}, {Howard}, {Evans}, {Fowler}, {Fullmer},
  {Hurt}, {Light}, {Kopan}, {Marsh}, {McCallon}, {Tam}, {Van Dyk}, \&
  {Wheelock}}]{Skrutskie06}
{Skrutskie}, M.~F., {Cutri}, R.~M., {Stiening}, R., {et~al.} 2006, \aj, 131,
  1163, \dodoi{10.1086/498708}

\bibitem[{Soares-Santos {et~al.}(2019)}]{Soares-Santos:2019irc}
Soares-Santos, M., {et~al.} 2019, Astrophys. J. Lett., 876, L7,
  \dodoi{10.3847/2041-8213/ab14f1}

\bibitem[{{St{\"o}lzner} {et~al.}(2018){St{\"o}lzner}, {Cuoco}, {Lesgourgues},
  \& {Bilicki}}]{Stolzner}
{St{\"o}lzner}, B., {Cuoco}, A., {Lesgourgues}, J., \& {Bilicki}, M. 2018,
  \prd, 97, 063506, \dodoi{10.1103/PhysRevD.97.063506}

\bibitem[{{Takahashi} {et~al.}(2012){Takahashi}, {Sato}, {Nishimichi},
  {Taruya}, \& {Oguri}}]{Takahashi12}
{Takahashi}, R., {Sato}, M., {Nishimichi}, T., {Taruya}, A., \& {Oguri}, M.
  2012, \apj, 761, 152, \dodoi{10.1088/0004-637X/761/2/152}

\bibitem[{Taylor {et~al.}(2012)Taylor, Gair, \& Mandel}]{Taylor:2011fs}
Taylor, S.~R., Gair, J.~R., \& Mandel, I. 2012, Phys. Rev. D, 85, 023535,
  \dodoi{10.1103/PhysRevD.85.023535}

\bibitem[{{van der Walt} {et~al.}(2011){van der Walt}, {Colbert}, \&
  {Varoquaux}}]{2011CSE....13b..22V}
{van der Walt}, S., {Colbert}, S.~C., \& {Varoquaux}, G. 2011, Computing in
  Science and Engineering, 13, 22, \dodoi{10.1109/MCSE.2011.37}

\bibitem[{Verde {et~al.}(2019)Verde, Treu, \& Riess}]{Verde:2019ivm}
Verde, L., Treu, T., \& Riess, A. 2019, in {Tensions between the Early and the
  Late Universe}, \dodoi{10.1038/s41550-019-0902-0}

\bibitem[{{Wright} {et~al.}(2010){Wright}, {Eisenhardt}, {Mainzer}, {Ressler},
  {Cutri}, {Jarrett}, {Kirkpatrick}, {Padgett}, {McMillan}, {Skrutskie},
  {Stanford}, {Cohen}, {Walker}, {Mather}, {Leisawitz}, {Gautier}, {McLean},
  {Benford}, {Lonsdale}, {Blain}, {Mendez}, {Irace}, {Duval}, {Liu}, {Royer},
  {Heinrichsen}, {Howard}, {Shannon}, {Kendall}, {Walsh}, {Larsen}, {Cardon},
  {Schick}, {Schwalm}, {Abid}, {Fabinsky}, {Naes}, \& {Tsai}}]{Wright10}
{Wright}, E.~L., {Eisenhardt}, P. R.~M., {Mainzer}, A.~K., {et~al.} 2010, \aj,
  140, 1868, \dodoi{10.1088/0004-6256/140/6/1868}

\bibitem[{{Xavier} {et~al.}(2019){Xavier}, {Costa-Duarte},
  {Balaguera-Antol{\'\i}nez}, \& {Bilicki}}]{Xavier19}
{Xavier}, H.~S., {Costa-Duarte}, M.~V., {Balaguera-Antol{\'\i}nez}, A., \&
  {Bilicki}, M. 2019, \jcap, 2019, 037, \dodoi{10.1088/1475-7516/2019/08/037}

\bibitem[{Zhou {et~al.}(2021)}]{Zhou:2020nwq}
Zhou, R., {et~al.} 2021, Mon. Not. Roy. Astron. Soc., 501, 3309,
  \dodoi{10.1093/mnras/staa3764}

\bibitem[{Zonca {et~al.}(2019)Zonca, Singer, Lenz, Reinecke, Rosset, Hivon, \&
  Gorski}]{Zonca2019}
Zonca, A., Singer, L., Lenz, D., {et~al.} 2019, Journal of Open Source
  Software, 4, 1298, \dodoi{10.21105/joss.01298}

\end{thebibliography}
\bibliographystyle{aasjournal}



\end{document}